\documentclass[a4paper]{article}

\usepackage{amssymb}
\usepackage{amsmath,amsthm}      % some advanced mathematics notation
\usepackage{graphicx}            % easy inclusion and manipulation of images
\usepackage{microtype}           % Just for fun, really hone in with the typography
\usepackage{booktabs}            % Really, really nice looking tables
\usepackage{multicol}            % Multiple columns in a local setting
\usepackage[hidelinks]{hyperref} % clickable references in the output, but hide them visually
\usepackage{physics}
\usepackage{mathtools}
\usepackage{relsize}
\usepackage{setspace}
\usepackage{gensymb}
\usepackage{placeins}

\usepackage{perpage} 
\usepackage{chngcntr}
\theoremstyle{definition}
\usepackage[titletoc,title]{appendix}
\usepackage{tocloft}
\usepackage[all]{xy}
\usepackage{sectsty}
\usepackage[top=25mm, bottom=25mm, left=15mm, right=15mm]{geometry}
\usepackage{authblk}
\usepackage[titletoc,title]{appendix}
\usepackage{tikz}
\usepackage{xcolor}
\usepackage{soul}
\usepackage{float}

\theoremstyle{definition}

\newcommand{\mathsym}[1]{{}}
\newcommand{\unicode}[1]{{}}
\newcommand{\ba}{\begin{align}}
\newcommand{\ea}{\end{align}}

\usepackage{natbib}
\setcitestyle{super,open={},close={},comma}

\title{General method for solving nonlinear optical scattering problems using fix point iterations}
\author[1]{Per Kristen Jakobsen}
\affil[1]{Department of Mathematics and Statistics, the Arctic University of Norway,\linebreak Troms\o, Norway, per.jakobsen@uit.no, }
\begin{document}

\maketitle

\begin{abstract}
In this paper we introduce a new fix point iteration scheme for solving nonlinear electromagnetic scattering problems. The method is based on  a spectral formulation of Maxwell's equations called the Bidirectional Pulse Propagation Equations. The scheme can be applied to a wide array of slab-like geometries, and for arbitrary material responses. We derive the scheme and  investigated how it performs with respect to convergence and accuracy by applying it to the case of light scattering from a simple slab whose nonlinear material response is a sum a very fast electronic vibrational response, and a much slower molecular vibrational response.
\end{abstract}

\tableofcontents

\section{Introduction}
Propagation problems in electromagnetics in the presence of linearly dispersive materials, has always been problematic. The key reason for this is that these kinds of problems cannot easily be solved as an initial value problem because the response of such materials are non-local in time. Over the years much work has been focused on resolving this problem. Two quite distinct approaches has been developing in parallel. 

In the first approach\cite{Yee,Tafloe} one makes use of the fact that the reason why the non local terms appear in Maxwell's equations is that we refrain from modeling the actual interaction between the electromagnetic field and the atoms comprising the material. If we chose to do this we would have to solve a coupled system consisting of Maxwells equations, describing the electromagnetic field,  and the Schrodinger equation describing the atoms. The resulting system is local in time. However, even today, a solution of propagation problems  along these lines, for example in the optical range, are beyond reach for macroscopic pieces of materials. The reason for this is the extremely wide time- and space scales involved in such a calculation. However, if the pulse to be propagated  is narrow spectrally, and the response of the dispersive material is fairly simple within the spectral range of the pulse, in the sense that there are few resonances, solving the pulse propagation as an initial value problem becomes feasible. One simply design a system of ordinary differential equations, driven by the electromagnetic field, which at each point in space has approximately the same response as the material within the spectral range of the pulse\cite{Hagness}. The resulting system of equations is now local in time and can be run as an initial value problem. This is the approach used in the well known and much used Finite Difference Time Domain(FDTD) method. For situations where the pulses are very broad spectrally and/or the material response is spectrally complex within the temporal spectral range of the pulse, this numerical approach to dispersive pulse propagation quickly becomes very challenging.

In the second approach, one solves the pulse propagation problem as a space propagation problem,  in the spectral domain. Traditionally, the propagation of the spectral amplitudes is assumed to proceed along the z-axis. The spectral amplitudes are functions of frequency and transverse wave numbers, and in order for the approach to work, these space-time spectral amplitude must be known at some chosen point $z=a$. The great advantage of this approach is that it can easily accommodate materials with arbitrary temporal response spectrum and pulses that are very broad and complex, spectrally. And currently, these are exactly the kind of situations that are of key interest. Another great advantage of this second approach, over the first approach,  is that for propagation over distances that are long compared to the center wavelength of the pulse, this second approach is much faster and more accurate than any method based on the first approach. It should however be noted that if the pulse contains a large fraction of spatially spectral components, plane waves, that travel at a significant angle away from the $z$-axis, much of the speed advantage may be lost. Currently, the best known such method is the Unidirectional Pulse Propagation Equation(UPPE)\cite{Miro1,Miro2,Miro4,Vegard,Per2}. All methods of this type, UPPE included, does however have to contend with a key problem; the approach requires the full temporal spectrum at some point in space, which means that one is required to know the pulse at this point in space, for all time, past, present, and future. This is in general not possible. If the material the pulse is traveling through contains material interfaces and/or regions of strong nonlinear response, reflected pulses are created. These reflected pulse travel back along the z-axis, and at some point returns to the point $z=a$. Thus, typically, the full temporal spectrum at the point $z=a$ contains elements that can only be known after the pulse has travelled through the material. Therefore, for many important situations the full temporal spectrum cannot be given at any point, and, thus, UPPE, or any method based on this second approach, fails to work. This said, it is worth noting that very frequently reflections from regions of nonlinear material response are very small and can be neglected\cite{Miro3}. Thus, this second approach, UPPE in particular, tackles most nonlinear material responses well.

More recently, a third approach to optical pulse propagation has been introduced. This is also a space propagation method for the spectral amplitude(s) of the pulse. Assuming as usual that space propagation is along the $z$-axis,  what distinguishes this third approach from the second approach, is that the full spectral amplitudes  are split into left and right moving spectral components and each of these separate components are propagating along the $z$-axis. This method retains one of the great advantages of the second approach, in that it can easily accommodate pulses with very broad temporal spectra and temporally complex  material response. The speed advantage of the second method with respect to the first method is however not maintained, and thus, using this approach for propagating pulses over distances very large compared to to the center wave length of the pulse, is not currently computationally feasible. The method using this third approach that is most similar to UPPE is the Bidirectional Pulse Propagation Equation(BPPE)\cite{Per1}. This method can handle all sorts of reflections whether from material interfaces or from nonlinear material response. The key problem with this third approach, which BPPE is designed to tackle, is the fact that the coupled $z$-propagation equations for the right and left moving spectral components cannot be solved because we know, and can control, only the right moving spectral component at any point $z=a$, assuming the source for the light pulse is on the left of $z=a$. The left moving component cannot be known at this initial point since it is generated by interfaces and/or nonlinearities to the right of $z=a$. This is in fact the exact same problem that doomed UPPE for problems with significant reflection. The way BPPE resolve this conundrum is through a reformulation into a fix point problem for a nonlinear map designed from the propagation equations for the spectral components and electromagnetic boundary conditions at material interfaces\cite{Andrew,Magdalena}. 

What is new in the current paper, as compared to previous papers on this problem, is the redesign of the nonlinear map, a redesign that remove redundant computational elements included in earlier incarnations of BPPE, and that naturally reduces the pulse propagation problem to finding zeroes to a nonlinear map. Finding zeroes for nonlinear maps is an old problem that has been studied  extensively in both pure and applied mathematics, and finding zeroes for  the BPPE map can rely on this large body of work and methods. Furthermore, because nonlinear material response frequently is fairly weak, like  in  optics for example, it turns out that finding the zeroes for the nonlinear map can in many cases be reduced to a fix point problem for the deviation between the zero for the full nonlinear map and a known  map. We believe that this fix point problem can for the most part be solved by simply iterating the map.  This fact can be  computationally highly significant, because simply iterating a nonliner map is, per step, computationally much cheaper than for example, quasi-Newton methods, which  are often used for finding zeroes of nonlinear maps. However, that said, quasi-Newton methods typically have a faster rate of convergence than simple function iteration when we start close to the solution. So, what method is the most efficient one for finding zeroes to our nonlinear map is not evident, but in this paper, which is already long enough,  we do not investigate this issue further.

In this paper our aim is to offer supporting evidence that our new formulation of BPPE is computationally feasible, and that the function iteration does converge to the fix point. We will do this by applying our constructions to the case of light scattering from a slab of material where the nonlinearity  is  a sum of a fast electronic vibrational response, and slow molecular vibrational response.

\section{Scattering from a nonlinear slab}
Our new method for solving electromagnetic scattering problems can handle very general material responses, and also a wide selection of slab-like geometries. For this paper, however, where our aim is to illustrate, derive, and numerically
test, our new fix point iteration scheme, we chose the simplest geometric setting possible. That of a simple slab. By this we mean that space if filled up with a material whose optical response is independent of $x$ and $y$ with respect  to a given $xyz$ Cartesian coordinate system. Furthermore,  optical properties of the material are constant in the regions $z<a$, $a<z<b$ and $z>b$. We will in this paper denote these three regions by {\it in front} of the slab, {\it inside} the slab and {\it behind} the slab.

Our basic model for light scattering is of course the Maxwell equations
\begin{align}
    \nabla \times \mathbf{E} + \partial_t \mathbf{B} &= 0,\nonumber \\
    \nabla \times \mathbf{H} - \partial_t \mathbf{D} &= 0,\nonumber \\
    \nabla \cdot \mathbf{B} &= 0, \nonumber\\
    \nabla \cdot \mathbf{D} &= 0.\label{Maxwell}
\end{align}

With respect to material response, we will start by assuming that the material is nonmagnetic. Note however, that the method will work equally well if the material has nontrivial magnetic properties, linear, or nonlinear. Here, however, where our aim is to illustrate the method in a simple, but nontrivial context, we assume that the material is non-magnetic
\begin{align}
    \mathbf{B} &= \mu_0 \mathbf{H}.\label{Nonmagnetic}
\end{align}
It is well known that in the macroscopic Maxwell equations (\ref{Maxwell}), the electric response, defined by the polarization vector $\mathbf{P}$, is related to the electric displacement field $ \mathbf{D}$, through the identity
\begin{align}
    \mathbf{D} &=\epsilon_0 \mathbf{E}+ \mathbf{P}.\label{Displacement}
\end{align}

In the simplified setting investigated in this paper we will also assume that all nonlinear material response is located inside the slab for $a<z<b$. Furthermore we will assume that there are optical sources, both for $z<a$, and for $z>b$. The sources are aligned with the z-axis, located far away from the slab, and set up such that the optical fields generated by these two sources sources are uniform in the transverse direction and have perpendicular incidence, at $z=a$ and $z=b$,  with respect to the orientation of the slab.

\begin{figure}[h!]
   \centering
   \includegraphics[width=0.8
\linewidth]{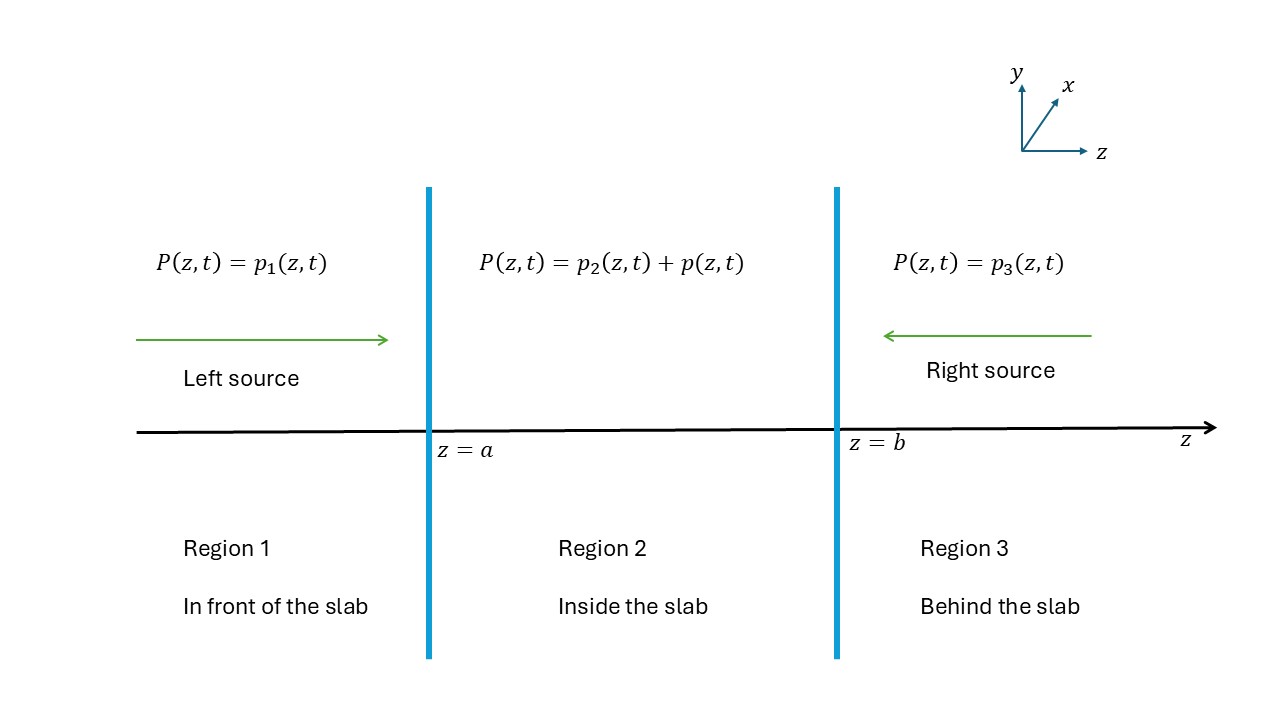}
    \caption{In this figure we display the geometry of the slab, the placement of the sources, and the material polarization response functions in each of the three regions defining the slab.}
    \label{slabGeometry}
\end{figure}

In figure $\ref{slabGeometry}$ we illustrate the slab geometry we are investigating in this paper.

Given the choice of slab geometry, and the assumed nature of the sources,  we may consistently assume that $\mathbf{E}$, $\mathbf{H}$ and $\mathbf{P}$, are of the form
\begin{align}
    \mathbf{E}(\mathbf{x}, t) &= e(z, t) \mathbf{j}, \nonumber\\
    \mathbf{H}(\mathbf{x}, t) &= h(z, t) \mathbf{i} + q(z, t) \mathbf{k}, \nonumber\\
    \mathbf{P}(\mathbf{x}, t) &= \epsilon_0 P(z, t) \mathbf{j},\label{SimplifiedFields}
\end{align}
where $\mathbf{i}$, $\mathbf{j}$, and $\mathbf{k}$, are the unit vectors along the axes of the $xyz$-Cartesian coordinate system.

For the scalar polarization $\epsilon_0 P(z,t)$, we assume that the material in front of the slab, $z<a$, inside the slab, $a<z<b$,  and behind the slab, $z>b$, has possibly different linear temporal dispersion, described by the polarization functions $\epsilon_0 p_1$, $\epsilon_0 p_2$ and $\epsilon_0 p_3$. Inside the slab we will in this section work with a completely general nonlinear polarization, $\epsilon_0 p(z,t)$. All this amounts to an expression for the function $P(z,t)$ of the form
\begin{align}
    P(z,t)&= p_1(z,t),\;\;z<a,\nonumber\\
    P(z,t) &= p_2(z,t)+p(z,t),\;\;a<z<b,\nonumber\\
   P(z,t) &= p_3(z,t),\;\;z>b,\label{polarizationModels1}
\end{align}
where the functions $p_i(z,t)$ are of the form
\begin{align}
    p_i(z, t) &= \int_{-\infty}^{t} ds \, \chi_i(t - s) e(z, s), \;\;i=1,2,3. \label{linearDispersion}
\end{align}

\subsection{The BPPE equations for a nonlinear slab}

Taking into account the assumptions made on the field geometry  and polarization in (\ref{Displacement}), (\ref{SimplifiedFields}), and (\ref{polarizationModels1}), we get the following simplified form of Maxwell equations  in the given slab geometry
\begin{align}
  -\partial_z e + \mu_0 \partial_t h &= 0,\nonumber \\
   \partial_z h - \epsilon_0 \partial_t (e + p_1) &= 0,\;\;z<a,\nonumber\\[4pt]
    -\partial_z e + \mu_0 \partial_t h &= 0,\nonumber \\
   \partial_z h - \epsilon_0 \partial_t (e + p_2) &= \epsilon_0\partial_t p,\;\;a<z<b,\nonumber\\[4pt]
    -\partial_z e + \mu_0 \partial_t h &= 0,\nonumber \\
   \partial_z h - \epsilon_0 \partial_t (e + p_3) &= 0,\;\;z>b.\label{simplifiedMaxwell}
\end{align}

In equations (\ref{simplifiedMaxwell}), we have disregarded the equation $\partial_z q(z,t)=0$, which is decoupled from the equations for the electric field and the transverse component of the magnetic field. The disregarded equation implies that any longitudinal component of the magnetic field can be time dependent, but must be uniform in space. Such a field plays no part in what we do in this paper, and is thus disregarded.

Recall that we allow for dual sources for the scattering problem, one source field approaching from the left, hitting the slab at $z=a$, and another source field approaching from the right, hitting the slab at $z=b$.

At this point it is convenient to scale time, space, magnetic field amplitude $h$, and electric field amplitude $e$ using the left source pulse. Thus we write
\begin{align}
t = \bar{t}t', \quad z = \bar{z}z', \quad e = \bar{e}e', \quad h = \bar{h}h',
\end{align}
where $\bar{z}$ is the center wavelength of the left source pulse, $\bar{t}$ is the cycle time of the center frequency of the left source pulse measured in Hz, and where $\bar{e}$ and $\bar{h}$ are determined by the center intensity, $\bar{I}$, of the left source pulse by the expressions
\begin{align}
\bar{e} = \sqrt{Z_0 \bar{I}}, \quad \bar{h} = \sqrt{\frac{\bar{I}}{Z_0}},\label{Scaling}
\end{align}
where $Z_0 = \sqrt{\frac{\mu_0}{\epsilon_0}}$. 

The scale for the nonlinear polarization, $\bar{p}$, where
\begin{align}
    p=\bar{p}p',\label{pScale}
\end{align}
is undetermined at this point. It will be fixed in section 3 where we consider a specific nonlinear response function.
Using the given scaling it is easy to verify that Maxwell's equations (\ref{simplifiedMaxwell}) take the form
\begin{align}
  -\partial_z e + \partial_t h &= 0,\nonumber \\
   \partial_z h -  \partial_t (e + p_1) &= 0,\;\;z<a,\nonumber\\[4pt]
    -\partial_z e +  \partial_t h &= 0,\nonumber \\
   \partial_z h -  \partial_t (e + p_2) &=\epsilon\partial_t p,\;\;a<z<b,\nonumber\\[4pt]
    -\partial_z e +  \partial_t h &= 0,\nonumber \\
   \partial_z h -  \partial_t (e + p_3) &= 0.\;\;z>b,\label{DimensionlessMaxwell}
\end{align}
Note that we have removed primes on all quantities since the scaled fields are the only ones appearing in the rest of the paper,  and we have also introduced the symbol 
\begin{align}
    \epsilon=\frac{\bar{p}}{\sqrt{\bar{I}}}\label{eps},
\end{align} 
for the sole dimensionless parameter in the equations.
Clearly, $\epsilon$ measures the dimensionless size of the nonlinear polarization inside the slab. The details regarding the derivation of the dimensionless system (\ref{DimensionlessMaxwell}) and the associated scales can be found in Appendix A.

In order to derive the BPPE equations in all the three separate domains for $z$, we need to express the electric and magnetic fields, $e$ and $h$, using the modes of the linear part of the Maxwell equations (\ref{DimensionlessMaxwell}), in each domain. These expressions are straight forward  to find, and are given by

\begin{align}
\begin{pmatrix}
e \\
h
\end{pmatrix}
(z, t) &= \int_{-\infty}^{\infty} d\omega \left\{ A_+(z, \omega) 
\begin{pmatrix}
\omega \\
-\beta_1
\end{pmatrix}
e^{i\beta_1 z} + A_-(z, \omega) 
\begin{pmatrix}
\omega \\
\beta_1
\end{pmatrix}
e^{-i\beta_1 z} \right\} e^{-i\omega t}, \quad z < a,\nonumber\\[4pt]
\begin{pmatrix}
e \\
h
\end{pmatrix}
(z, t) &= \int_{-\infty}^{\infty} d\omega \left\{ A_+(z, \omega) 
\begin{pmatrix}
\omega \\
-\beta_2
\end{pmatrix}
e^{i\beta_2 z} + A_-(z, \omega) 
\begin{pmatrix}
\omega \\
\beta_2
\end{pmatrix}
e^{-i\beta_2 z} \right\} e^{-i\omega t}, \quad a < z < b,\nonumber\\[4pt]
\begin{pmatrix}
e \\
h
\end{pmatrix}
(z, t) &= \int_{-\infty}^{\infty} d\omega \left\{ A_+(z, \omega) 
\begin{pmatrix}
\omega \\
-\beta_3
\end{pmatrix}
e^{i\beta_3 z} + A_-(z, \omega) 
\begin{pmatrix}
\omega \\
\beta_3
\end{pmatrix}
e^{-i\beta_3 z} \right\} e^{-i\omega t}, \quad z > b,\label{ModeExpansions}
\end{align}
where the propagation constants $\beta_i(\omega)$ in (\ref{ModeExpansions}) are given by the formulas
\begin{align}
\beta_i(\omega) = |\omega| n_i(\omega), \quad n_i(\omega) = \sqrt{1 + \frac{\hat{\chi}_i(\omega)}{\sqrt{2\pi}}}.\label{ComplexSquareRoot}
\end{align}

Note that, in general, $\hat{\chi}_i(\omega)$ is a complex quantity, and that the square root in (\ref{ComplexSquareRoot}) is the one which is positive for positive real numbers. Also note that the appearance of the factor $\sqrt{2\pi}$, in the relation between the Fourier transform of the function $\chi_i(t)$ from (\ref{linearDispersion}), $\hat{\chi}_i(\omega)$, and the index of refraction, occurs because we in this paper are using a symmetric version of the Fourier transform given by the formulas
\begin{align}
    \hat{f}(\omega)&=\frac{1}{\sqrt{2\pi}}\int_{-\infty}^{\infty}dt\;f(t)e^{i\omega t},\nonumber\\
    f(t)&=\frac{1}{\sqrt{2\pi}}\int_{-\infty}^{\infty}d\omega\;\hat{f}(\omega)e^{-i\omega t}.\label{MyFourier}
\end{align}

If we did not allow for the spectral amplitudes $A_+(z, \omega)$, and $A_-(z, \omega)$ to depend on the $z$-variable , the expansions (\ref{ModeExpansions}) would be the exact solution to Maxwell's equations (\ref{DimensionlessMaxwell}), outside and inside the slab, for the linear case, where the function $p(z,t)$,  is equal to zero. 

When we allow for a $z$-dependence of the spectral amplitudes, the expansions given can be used to express any two-component vector field $(v_1(z,t),v_2(z,t))$, whatsoever, in particular the two-component vector field $(e(z,t),h(z,t))$, solving Maxwells equations outside, and inside, the slab when the function $p(z,t)$, is nonzero. This statement follows from the completeness of the modes comprising the mode expansions (\ref{ModeExpansions}).

It is easy to verify that $e$ and $h$ in formulas (\ref{ModeExpansions}) are real only if the constraint 
\begin{align}
    A_+(z, \omega)=-A_-(z, -\omega)^*,\label{RealityConditions}
\end{align}
is imposed on the spectral amplitudes. Here, for any complex quantity $z$, the complex conjugate is denoted by $z^*$.

Recall that these relations implies for example that the two spectral amplitudes $ A_+(z, \omega), A_-(z, \omega)$ are fully specified if they are given for positive frequencies only. Note also that, in particular, these relations also implies that the mode formula for the electric field in the range $a<z<b$ can be written in the form
\begin{align}
e(z, t) &= \int_{0}^{\infty} d\omega \left\{ \omega A_+(z, \omega) 
e^{i\beta_2 z} + \omega A_-(z, \omega) 
e^{-i\beta_2 z} \right\} e^{-i\omega t} + (*),\label{eFieldCausalityForm}
\end{align}
where for any complex quantity $z$, $z+(*)= z+z^*$. Similar formulas holds for the electric field in front of the slab, $\beta_2\rightarrow\beta_1$, and behind the slab, $\beta_2\rightarrow\beta_3$.
These formulas for the electric field  will be used later in this paper when we discuss causality for our field constructions.

Inserting the expansions (\ref{ModeExpansions}) into Maxwell's equations (\ref{DimensionlessMaxwell}), taking into account that the terms in the expansions are solutions to (\ref{DimensionlessMaxwell}) for the case when $p=0$, we find, after a small amount of algebra, that the following equations for the spectral amplitudes outside, and inside, the slab, holds

\begin{align}
    \partial_z A_+(z, \omega) &= 0, \nonumber\\
    \partial_z A_-(z, \omega) &= 0, \quad z < a,\nonumber\\[6pt]
    \partial_z A_+(z, \omega) &= i\epsilon\frac{\omega e^{-i\beta_2 z}}{2\beta_2} \hat{p}(z, \omega), \nonumber\\
    \partial_z A_-(z, \omega) &= -i\epsilon\frac{\omega e^{i\beta_2 z}}{2\beta_2} \hat{p}(z, \omega), \quad a < z < b,\label{BPPE}\\[6pt]
    \partial_z A_+(z, \omega) &= 0,\nonumber \\
    \partial_z A_-(z, \omega) &= 0, \quad z > b.\nonumber
\end{align}
These are the BPPE equations for the given slab geometry and any nonlinear polarization.
Note that in these equations, $\hat{p}(z,\omega)$  is the temporal Fourier transform of the function $p(z,t)$. Note also that the condensed notation used to express the BPPE equations might lead one to think that the differential equations inside the slab are uncoupled with respect to $\omega$. This is true only for the case of a purely linear optical response for which  $\hat{p}(z,\omega)$ is a linear function of the spectral amplitudes $A_+(z, \omega)$, and $A_-(z, \omega)$. In general, the function $\hat{p}(z,\omega)$ is a nonlinear {\it functional} depending on the spectral amplitudes over their whole spectral range.

\subsection{Transforming the BPPE equation into a nonlinear mapping problem.}\label{interpretation}
Let us start by introducing a useful interpretation of the spectral amplitudes $A_+(z, \omega)$, and $A_-(z, \omega)$. From the expansions (\ref{ModeExpansions}) it appears that $A_+(z, \omega)$ is the spectral amplitude for right-moving waves, and that $A_-(z, \omega)$ is the spectral amplitude for left-moving waves. For situation where the electromagnetic pulses are traveling freely, without any significant reflection, this interpretation is always correct. However, for situations where we do have a significant amount of reflection, like in this paper, the interpretation may not always be correct. We will return to this issue later in the paper.

Note that the solution to the BPPE equations inside the slab, $a < z < b$, defines a map, $U(a,b)$, of spectral amplitudes
\begin{align}
\begin{pmatrix}
A_+ \\
A_-
\end{pmatrix}
(a, \omega) \xrightarrow{U(a, b)}
\begin{pmatrix}
A_+ \\
A_-
\end{pmatrix}
(b, \omega).
\end{align}
As noted at the end of the previous section, the condensed notation we are using could lead one to think that $U(a,b)$ maps the spectral amplitudes at $a$, for each $\omega=\bar{\omega}$ to the spectral amplitudes at $b$, at the {\it same} frequency $\omega=\bar{\omega}$. This is not the case, in general 
\begin{equation}
    U(a,b)\begin{pmatrix}
A_+ \\
A_-
\end{pmatrix}(a,\omega),
\end{equation}
depends on the spectral amplitudes $A_+(a, \omega)$, and $A_-(a, \omega)$ over their whole spectral range.

It is clear from our constructions that any solution $A_+(z, \omega) , A_-(z, \omega)$, to the BPPE equations, where $z$ ranges over the open intervals $(-\infty,a)$, $(a,b)$ and $(b,\infty)$, ensure that $e(z,t)$, $h(z,t)$ satisfy (\ref{DimensionlessMaxwell}), and consequently that $\mathbf{E}$, and $ \mathbf{H}$, from (\ref{SimplifiedFields}), together with  $\mathbf{D}$, and $\mathbf{B}$, from  (\ref{Nonmagnetic}) and (\ref{Displacement}), satisfy Maxwells equations (\ref{Maxwell}), in the same three open intervals.

At the two points $z=a,b$, the electromagnetic field has to satisfy the well known electromagnetic boundary conditions. Since these conditions are linear in the fields, and since the boundary points are fixed in time, it is clear that the electromagnetic boundary conditions turns into linear maps, $M_{12}(\omega)$, and $M_{23}(\omega)$, mapping the spectral amplitudes at $z=a^-$ to the corresponding spectral amplitudes at $z=a^+$, and the spectral amplitudes at $z=b^-$ to the ones at $z=b^+$. Here $a^-$ is a notation for the limit of $z$, when $z$ approach $a$ from below, etc.  

The matrices $M_{12}(\omega)$ and $M_{23}(\omega)$ are easy to construct, and are given by 

\begin{align}
    M_{12}(\omega)&=M_2(a,\omega)^{-1}M_1(a,\omega),\nonumber\\
    M_{23}(\omega)&=M_3(b,\omega)^{-1}M_2(b,\omega),
\end{align}
where
\begin{align}
    M_i(z,\omega)&=\begin{pmatrix}
 e^{i\beta_i z} &  e^{-i\beta_i z} \\
-\beta_i e^{i\beta_i z}  &  \beta_i e^{-i\beta_i z}
\end{pmatrix},\quad i=1,2,3,\quad z=a,b
\end{align}

\begin{figure}[h!]
    \centering
    \includegraphics[width=0.8\linewidth]{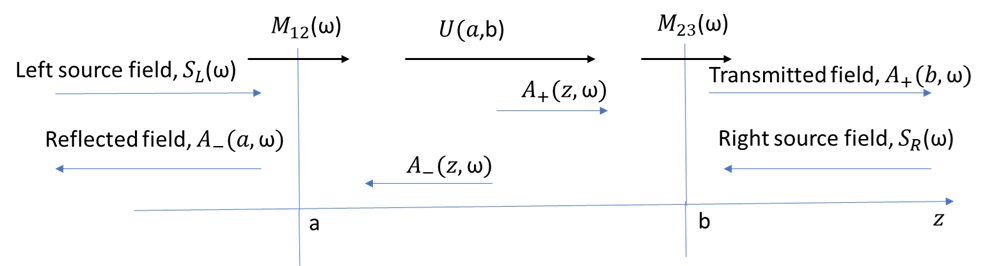}
    \caption{The nonlinear scattering problem displayed  using maps and spectral amplitudes defined in the text}
    \label{fig1}
\end{figure}

Denoting  the spectral amplitudes of the left and right sources by $S_L(\omega)$ and $S_R(\omega)$, the scattering problem for the nonlinear slab is illustrated in figure \ref{fig1}.

Note that in order to compute the spectral amplitudes of the electromagnetic field inside the slab, the first order BPPE equations (\ref{BPPE}) must be supplied with the spectral amplitude of the left source field, and for the  reflected field at $z=a$. The left source is given, but the reflected field cannot be given at $z=a$ since it depends on the nonlinear dynamics inside the slab, and also on the given right source at $z=b$.

 For this reason we define a map $G$ by the composition
\begin{align}
G(u) = P \circ M_{23}(\omega) \circ U(a, b) \circ M_{12}(\omega) 
\begin{pmatrix}
S_L(\omega) \\
u
\end{pmatrix}
\quad \text{Where } P 
\begin{pmatrix}
u \\
v
\end{pmatrix}
= v.
\end{align}
Solving the scattering problem for the nonlinear slab has now been reduced to solving the following equation
\begin{align}
G(A_-(a,\omega)) = S_R(\omega).\label{GMappingProblem}
\end{align}
Since the solution to this equation is the spectral amplitude of the reflected field at $z=a$, we can now use the BPPE equations to compute the spectral amplitudes $A_+(z, \omega) , A_-(z, \omega)$ inside the slab, and also the spectral amplitude $A_+(b, \omega)$ for the transmitted field behind the slab. From these amplitudes the electromagnetic field can be computed everywhere. 

Numerical solution of nonlinear mapping problems like (\ref{GMappingProblem}), has been investigating extensively over the years, and a large set of numerical algorithms are available. However, this said, for complex temporal spectra, whose resolution require a very large number of discrete spectral points, the map $G$ is defined on a very high dimensional space. This can make  the direct numerical solution of  (\ref{GMappingProblem}) costly. In the particular setting we are using in this paper, where there are one spatial, and one temporal dimension, the solution can be found using a reasonably high powered PC. However, for more spatial dimensions, a more powerful computational platform will be needed.

\subsection{Finding the zeroes of the nonlinear map using fix point iterations.}

In optical applications, the nonlinear material effects are very frequently dominated by the linear ones. This fact can be used to simplify the solution of the basic equation (\ref{GMappingProblem}). This is true because through the expansions (\ref{ModeExpansions}), the BPPE equations for the spectral amplitudes inside the slab is driven solely  by the nonlinear material response, whose size is fixed by the parameter $\epsilon$, defined in (\ref{eps}), a parameter which is frequently small. 

The smallness of $\epsilon$ means that the nonlinear map $U(a, b)$ is close to the identity map. Inspired by this, let us introduce the map $G_0$, defined by
\begin{align}
G_0(u) = P \circ M_{23} (\omega)\circ I \circ M_{12}(\omega)\begin{pmatrix}
S_L(\omega) \\
u
\end{pmatrix}.
\end{align}
Then, if $A_0(\omega)$ is the reflection spectrum for the linear scattering problem, we have
\begin{align}
G_0(A_0(\omega)) = S_R(\omega).
\end{align}
Let $a_- (\omega) \equiv A_-(a, \omega) - A_0(\omega)$. Thus, $a_- (\omega)$ is the deviation of the nonlinear reflection spectrum from the linear one.
Next define the nonlinear map $V$ by $V = U(a, b) - I$. Thus, $V$ is the deviation of the nonlinear map $U(a, b)$ from the identity map. 

Note that in the following calculations we will, for notational simplicity, stop listing the frequency dependence of any of the involved quantities. 

From the given definitions we then have
\begin{align}
U(a, b)&= I + V\nonumber,\\
A_-(a)&= A_0 + a_-.\nonumber
\end{align}
Given these definitions, we can rewrite the basic equation (\ref{GMappingProblem}) into the form
\begin{align}
    G(A_0 + a_-) &= P \circ M_{23} \circ U(a, b) \circ M_{12}  \begin{pmatrix} S_L \\ A_0 + a_- \end{pmatrix} \nonumber \\
    &= P \circ M_{23} \circ (I + V) \circ M_{12}  \begin{pmatrix} S_L \\ A_0 + a_- \end{pmatrix} \nonumber \\
    &= P \circ M_{23} \circ I \circ M_{12} \begin{pmatrix} S_L \\ A_0 \end{pmatrix}  \nonumber\\
    &\quad + P \circ M_{23} \circ I \circ M_{12}  \begin{pmatrix} 0 \\ a_- \end{pmatrix} \nonumber \\
    &\quad + P \circ M_{23} \circ V \circ M_{12}  \begin{pmatrix} S_L \\ A_0 + a_- \end{pmatrix}\nonumber  \\
    &= G_0(A_0) \nonumber\\
    &\quad + P \circ M_{23} \circ I \circ M_{12}  \begin{pmatrix} 0 \\ a_- \end{pmatrix} \nonumber\\
    &\quad + P \circ M_{23} \circ V \circ M_{12} \begin{pmatrix} S_L \\ A_0 + a_- \end{pmatrix}  = S_R.\label{GReduction1}
    \end{align}
Using the fact that 
\begin{align}
\begin{pmatrix} S_L \\ A_0 + a_- \end{pmatrix}&= \begin{pmatrix} S_L \\ A_0 \end{pmatrix} + \begin{pmatrix} 0 \\ a_- \end{pmatrix},\nonumber\\
\begin{pmatrix} 0 \\ a_- \end{pmatrix}&= \begin{pmatrix} 0 \\ 1 \end{pmatrix} a_-,\nonumber
\end{align}
equation (\ref{GReduction1}) can be rewritten into the form
\begin{align*}
    &P \circ M_{23} \circ I \circ M_{12} \begin{pmatrix} 0 \\ 1 \end{pmatrix} a_-\nonumber \\
    &\quad + P \circ M_{23} \circ V \circ M_{12} \begin{pmatrix} S_L \\ A_0 + a_- \end{pmatrix} = 0.
\end{align*}

Defining the complex quantity $\gamma(\omega) \neq 0$ by the expression
\begin{align*}
\gamma \overset{\text{def}}{=} P \circ M_{23} \circ I \circ M_{12} \begin{pmatrix} 0 \\ 1 \end{pmatrix} ,
\end{align*}
and the nonlinear map $H(u)$ by the expression
\begin{align*}
H(u)\equiv -\left( \frac{1}{\gamma} \right) P \circ M_{23} \circ V \circ M_{12} \begin{pmatrix} S_L \\ A_0 + u \end{pmatrix},
\end{align*}
our basic equation (\ref{GMappingProblem}) takes the form
\begin{align*}
a_- = H(a_-)
\end{align*}
We propose to solve this fix point problem by a simple iteration
\begin{align}
a_-^{n+1} = H(a_-^n).\label{FixPointIteration}
\end{align}

\subsection{A purely nonlinear slab}
In our numerical tests we will exclusively be focused on the case of a purely nonlinear slab, by which we mean a situation where 
\begin{align*}
    \beta_i &= |\omega|, \quad i = 1,2,3 \\
    &\Downarrow \\
    M_{12} &= M_{23} = I \\
    &\Downarrow \\
    \gamma &= 1,\quad H(u) = P \circ V  \begin{pmatrix} S_L \\ A_0 + u \end{pmatrix}.
\end{align*}
This particular case is probably not easy to realize physically. But that does not really matter here, since our aim at this point is to test the method itself, in a situation that is nontrivial, but otherwise as simple as possible.
For this case, the iteration takes the simple form
\begin{equation}
    a^{n+1} = P \circ V \begin{pmatrix} S_L \\ A_0 + a^{n} \end{pmatrix}.\label{FixpointIteration}
\end{equation}

\section{Numerical simulations for a purely nonlinear slab}
For the numerical simulations we need to make a specific choice for the function determining the nonlinear polarization inside the slab, $p(z,t)$, which up to this point has been arbitrary. At this point we will assume that the nonlinear polarization  is  a sum of a Kerr-like fast electronic vibrational response,  $p_K$, and a Raman-like slow molecular vibrational response, $p_R$,

In dimensionless variables  this nonlinear polarization model takes the form

\begin{align}
    p(z,t)&=\left((1 - \theta) p_K(z,t) + \theta p_R(z,t)\right).\label{NonlinearPolarization}
\end{align}
Here  $\theta$ is a dimensionless constant measuring the relative size of the two components $p_K$ and $p_R$.

\begin{align}
    p_K(z, t) &=  e(z, t) \int_{-\infty}^{t} ds \, \chi_K(t - s) e^2(z, s), \\
    p_R(z, t) &= e(z, t) \int_{-\infty}^{t} ds \, \chi_R(t - s) e^2(z, s).
\end{align}
The scale for the nonlinear polarization is in Appendix A found to be of the form $\bar{p}=\eta\left(Z_0\bar{I}\right)^{\frac{3}{2}}$, were $\eta$ is a dimensional parameter which measures the  size of the nonlinear polarization, as compared to the linear one.

From this, and (\ref{eps}), we find that the sole dimensionless parameter in the model equations (\ref{DimensionlessMaxwell}) is given by $\epsilon=\eta Z_0\bar{I}$.
For the spectra of the two nonlinear dispersion functions we assume the form
\begin{align}
    \hat{\chi}_K(\omega) &= \frac{\Omega_e^2}{\Omega_e^2 - 2i\gamma_e \omega - \omega^2}, \\
    \hat{\chi}_R(\omega) &= \frac{\Omega_m^2}{\Omega_m^2 - 2i\gamma_m \omega - \omega^2}.\label{NonlinearDispersionFunctions}
\end{align}
All parameters in these expressions are of course scaled with respect to the center frequency of the left source, as discussed earlier in this paper, and in Appendix A.

In our numerical simulations we use the following values for the dimensionless quantities
\[
\begin{aligned}
    \epsilon &= 0.0067, \\
     \theta &= 0.5, \\
    \Omega_e &= 0.64, \\
    \gamma_e &= 3.65, \\
    \Omega_m &= 0.64, \\
    \gamma_m &= 0.23.
\end{aligned}
\]
These numerical values correspond to time constants for the electronic and molecular nonlinear response of 2 fs and 32 fs. The center wavelength and peak intensity of the left source is chosen to be 2.19 $\mu$m and $1.78 \times 10^{15} \, \frac{W}{m^2}$.

In all numerical calculations in this paper, the right source is set to zero, and the left source, which is acting at the left end of the slab, $z=a$, is taken to be
\begin{align}
    e(a,t)=\int_{-\infty}^{\infty}d\omega\;\left(\omega A_+(a,t)e^{i|\omega| a}+\omega A_-(a,t)e^{-i|\omega| a}\right)e^{-i\omega t},\label{eLeftSource}
\end{align}
where
\begin{align}
    A_+(a,\omega)&=\begin{cases}\; \; a_0e^{-\gamma(\omega-1)^2}e^{i\omega a},\;\; \omega>0\\
\; \; 0\quad\quad\quad\quad\quad,\;\;\omega <0\end{cases} \nonumber\\
\nonumber\\
A_-(a,\omega)&=\begin{cases}\; \; 0\quad\quad\quad\quad\quad\quad,\;\;\omega >0\\
\; \; -a_0e^{-\gamma(\omega+1)^2}e^{i\omega a},\;\;\;\omega<0\end{cases} .\label{ALeftSource}
\end{align}
Note that these expressions for $A_+$, and $A_-$ satisfy the reality condition (\ref{RealityConditions}), which they must.
In our numerical calculations we use the value $\gamma=50$, and for $a_0$ the value $0.2$. The dimensionless parameter $a_0$ is a scaling factor which ensure that the maximum of the absolute value of the electric field (\ref{eLeftSource}), is equal to unity.
For the purely nonlinear slab which we are considering in this section, there is translational invariance, so, here, we will assume without loss of generality that $a=0$.

\begin{figure}[h!]
    \centering
    \includegraphics[width=0.6\linewidth]{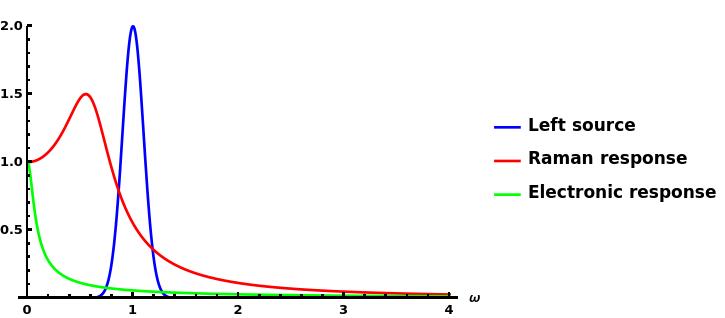}
    \caption{The amplitudes of Raman and electronic spectral response functions, $\hat{\chi}_R(\omega)$, and $\hat{\chi}_K(\omega)$, together with norm of spectral amplitude of the left source $A_+(a,\omega)$, scaled for convenience by factor of $0.4$.  Because of translational invariance we assume, without loss of generality, that $a=0$. }
    \label{fig2}
\end{figure}

In Figure \ref{fig2} we plot the two nonlinear dispersion functions (\ref{NonlinearDispersionFunctions}), together with the left source, for $0<\omega <4 $.

\FloatBarrier

In the next two subsections, we will not evaluate the accuracy of our method by comparing it to the exact solution, since the exact solution is not known. We will return to this kind of evaluation in a later section in this paper.  What we will do in the two following subsections is to look into the convergence of the fix point iteration method (\ref{FixPointIteration}), and determine to which accuracy the prospective fix point solution satisfy the exact equation (\ref{GMappingProblem}) for the nonlinear scattering problem.  Based on the prospective fix point solution we will also construct space-time density plots of the right moving and left moving parts of the electric field in front of, inside, and behind the slab. The question we ask here is if these space-time plots make physical sense, in particular if the constructed solutions are causal. Maxwell's equations are of course causal, but for the way we construct approximate solutions, using fix point iterations, derived from equation (\ref{GMappingProblem}),  the issue of causality is hidden. It is simply not obvious that the iteration method will converge to a solution that respects causality. From the construction of equation (\ref{GMappingProblem}), we {\it do} know that any solution to the electromagnetic scattering problem for our slab {\it will} be a solution of equation (\ref{GMappingProblem}). 

What is not obvious is whether or not this equation can have other solutions, perhaps solutions that do not respect causality, and if our iteration  method could end up converging to one of these solutions.

\subsection{A short slab}
In this sub section we will consider the case of a slab of length equal to $50$ wavelengths. For the scaling we are using that corresponds to a slab of length approximately equal to $110 \mu m$.

\begin{figure}[h!]
    \centering
    \includegraphics[width=0.6\linewidth]{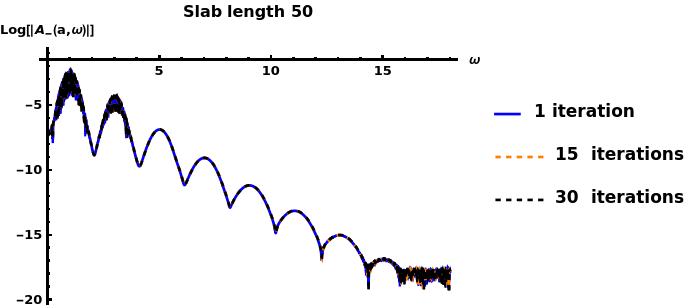}
    \caption{Logarithmic amplitude spectrum for the reflected field of a slab of length 50 wavelengths. The shape of the spectrum for iteration number 1, 15, and 30, for the fix point iterations of the map (\ref{FixpointIteration}), are included as solid blue, dashed yellow and dashed black curves. Zero padding $\omega_m=18$ }
    \label{fig3}
\end{figure}

In Figure \ref{fig3} we start with $a_-(\omega)$ equal to the zero function, which corresponds to an initial  guess for the reflection spectrum given by $A_-(0,\omega)=S_R(\omega)=0$, where we recall that, by assumption,  there is no source hitting the rear end of the slab. We show the first, fifteenth, and the thirtieth iteration. The iteration certainly seems to have converged, in fact all three iterates are indistinguishable in the plot. 

However, the approximate reflection spectrum displayed in Figure \ref{fig3} does not appear to be resolved, there are some very fast oscillations on top of features that clearly are unresolved.

In the next figure \ref{fig4}, we address this issue by showing a blowup of the spectrum for $0<\omega<2$. We are using 4000 discrete frequency point in the finite frequency interval (-63, 63), corresponding to a frequency grid size $\triangle\omega\approx 0.016$. From this size of $\triangle\omega$, and figure \ref{fig4}, we conclude that the spectrum is indeed resolved.

\begin{figure}[h!]
    \centering
    \includegraphics[width=0.6\linewidth]{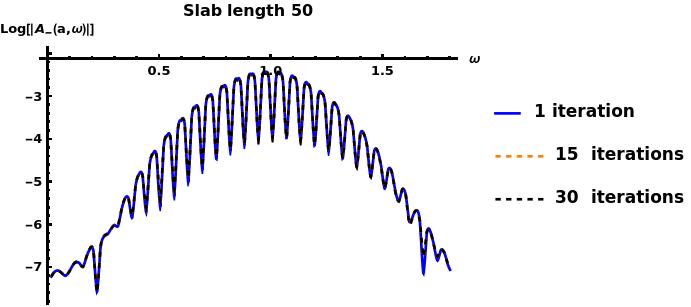}
    \caption{This is  the spectrum from Figure \ref{fig3} on the restricted interval $0<\omega<1.8$. Zero padding is $\omega_m=18$}
    \label{fig4}
\end{figure}
\begin{figure}[h!]
    \centering
    \includegraphics[width=0.6\linewidth]{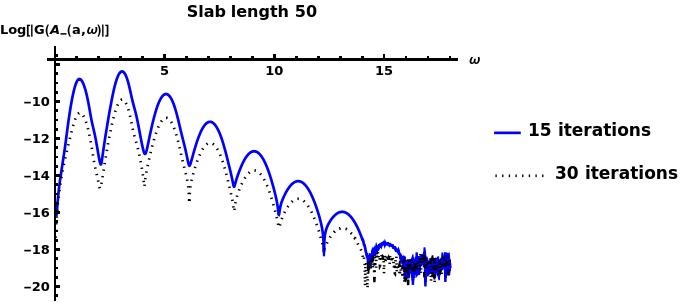}
    \caption{Logarithm of the remainder, $G(A_-(a))$, after 15 iterations, solid blue, and 20 iterations, dotted black, of the map (\ref{FixpointIteration}) for  a slab of length 50 wavelengths.  Zero padding is $\omega_m=18$ }
    \label{fig5}
\end{figure}

Before we proceed, there is one purely computational issue that needs to be addressed. The key computational step while using the iteration method described in this paper, is to evaluate the map $U(a,b)$, by  solve the BPPE equations inside the slab. The BPPE equations are differential equations in the spectral domain, and a well known problem with such equations is that they tend to amplify numerical noise existing in spectral ranges far from any relevant physical features of the spectrum. This problem is for example prevalent for UPPE, a much used, and thoroughly tested,  method for modeling high power laser pulse propagation in complex materials with very general optical response.\cite{Miro2,Miro1,Miro3}. The way the problem is resolved in UPPE is to set to zero all spectral amplitudes beyond a preselected zero-padding frequency value, $\omega_m$, at each step in the integration of the equations. Finding the right value for $\omega_m$, for any particular situation, typically require some  physical insight, combined with a certain amount of trial and error. We use the same approach here. For the calculations displayed in figures \ref{fig3},  \ref{fig4} and \ref{fig5},  we used $\omega_m=18$.

Spectral propagators like UPPE and BPPE are not unique in including an adjustable parameter like $\omega_m$. All discrete numerical methods approximating continuous ordinary  and/or partial differential equations have adjustable parameters whose value must be determined by a combination of insight and some trial and error. Choice of step length for discrete approximations to ODEs is an obvious example.

In figure \ref{fig5} we show a logarithmic plot of the residual of equation (\ref{GMappingProblem}) for $15$ and $30$ iteration. This figure tells us that the numerical  expression found after 30 iteration satisfy the exact equation (\ref{GMappingProblem}) to ten digits of accuracy. This certainly strengthen our belief that the iteration method has given us  a high precision numerical solution to light scattering from a purely nonlinear slab.  However, it is still worthwhile to verify that the found numerical solution is physical, in the sense that it respects causality.

For the case of a purely nonlinear slab, the general formula for the electric field in front of, inside, and behind the slab can according to (\ref{eFieldCausalityForm}) be written in the form 
\begin{align}
    e(z,t)&=\int_{0}^{\infty}d\omega\;\left(\omega A_+(z,\omega)e^{i\omega\left( z- t\right)}+\omega A_-(z,\omega)e^{-i\omega \left(z+ t\right)}\right)+ (*).\nonumber\\
\end{align}
Based on the nature of the modes in this expression, we split the electric field into a left moving field, $e_L(z,t)$, and a right moving field, $e_R(z,t)$ using the formulas
\begin{align}
    e_R(z,t)&=\int_{0}^{\infty}d\omega\;\left(\omega A_+(z,\omega)e^{i\omega\left( z-t\right)}\right) +(*),\nonumber\\
    e_L(z,t)&=\int_{0}^{\infty}d\omega\;\left(\omega A_-(z,\omega)e^{-i\omega\left( z+ t\right)}\right)+(*).\nonumber\\
\end{align}

\begin{figure}[h!]
    \centering
    \includegraphics[width=0.4\linewidth]{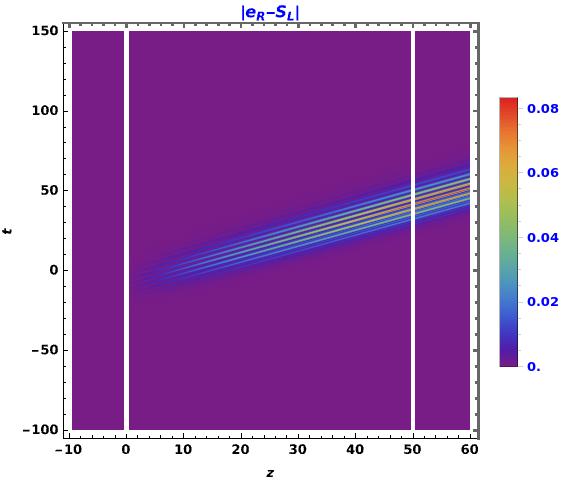}
    \caption{Space-time picture of the amplitude of the right moving part of the electric field inside a slab of length 50 wavelengths,  with the left source subtracted.   Zero padding is $\omega_m=18$ }
    \label{fig6}
\end{figure}

In figure \ref{fig6} we show the absolute value of the right moving part of the electric field. In the figure we have subtracted the left source since this source is much stronger than any right propagating field generated by nonliner interactions inside the slab. The direct source contribution  would thus completely dominate the picture, and a right propagating component, induced by nonlinear interactions, would be invisible. The right propagating field induced by nonlinear interactions of the source with the material, $e_R^{NL}(z,t)$ is defined on the spectral level through the formula
\begin{align}
    e_R^{NL}(z,t)&=\int_{0}^{\infty}d\omega\;\omega\left(A_+(z,\omega)-S_L(\omega)\right)e^{i\omega\left( z-t\right)}+(*).\nonumber
\end{align}

\begin{figure}[h!]
    \centering
    \includegraphics[width=0.4\linewidth]{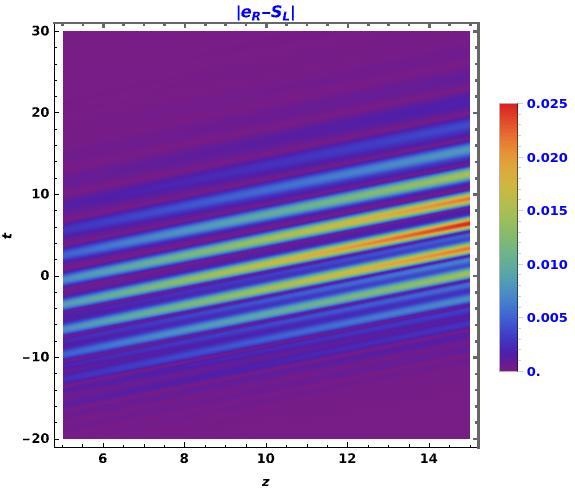}
    \caption{The amplitude of the right moving part of the electric field from figure \ref{fig6} restricted to the space-time box $(5,15)\cross(-20,30)$.}
    \label{fig65}
\end{figure}

This space-time picture of the right moving field $e_R^{NL}(z,t)$ makes perfect sense and is what one should expect. It shows a causal physical process where the source field, through nonlinear interactions with the underlying material in the slab, induces a gradual buildup of a right moving field inside the slab.

\begin{figure}[H]
    \centering
    \includegraphics[width=0.5\linewidth]{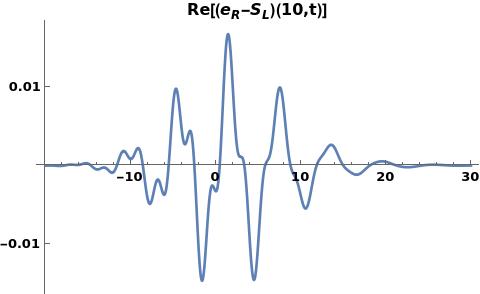}
    \caption{The right moving electric field as a function of time, taken at $z=10$, the center of the space-time box from figure \ref{fig65}}
    \label{fig7}
\end{figure}

In figure \ref{fig65} we zoom in on the space time box $(5,15)\cross(-20,30)$ from figure \ref{fig6}, and in figure \ref{fig7} we show the temporal structure of the induced right moving field at the center of the space time box, $z=10$. The pulse shows a clear temporal asymmetry. 

\begin{figure}[H]
    \centering
    \includegraphics[width=0.4\linewidth]{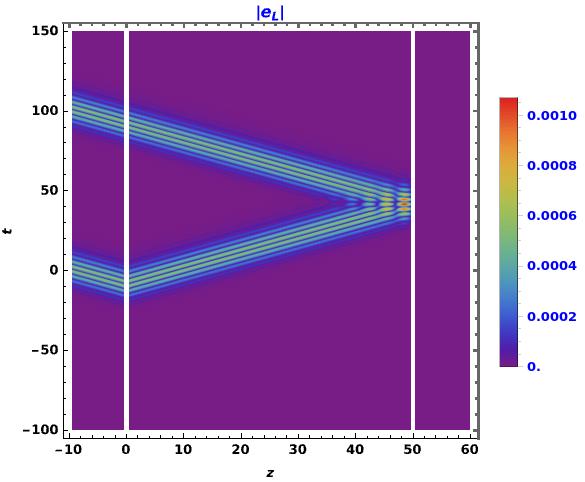}
    \caption{Space-time picture of the amplitude of the left moving part of the electric field. The white vertical lines defines the extent of a slab of length 50 wavelengths. Zero padding is $\omega_m=18$}
    \label{fig8}
\end{figure}

\begin{figure}[H]
    \centering
    \includegraphics[width=0.4\linewidth]{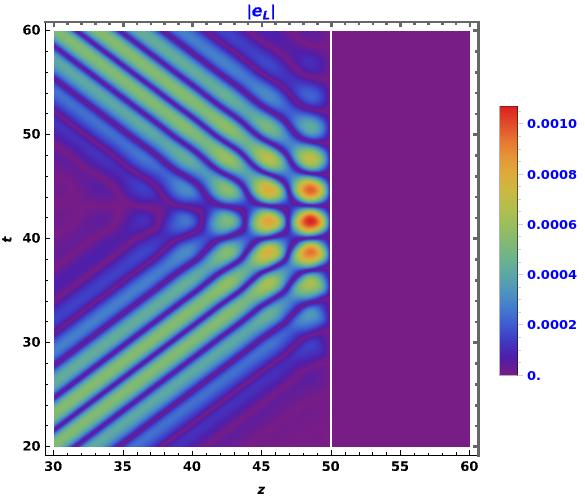}
    \caption{The amplitude of the left moving part of the electric field from figure \ref{fig8} restricted to the space-time box $(30,60)\cross(20,60)$.}
    \label{fig85}
\end{figure}

In figure \ref{fig8} we show a space-time picture of the left moving part of the electric field, and in figure \ref{fig85} we how a blowup of this field in a space-time box close to the right boundary of the slab.
At first glance this picture looks like a very familiar causal physical situation where a electromagnetic pulse reflects of the rear face of the slab. Since the slab is actually a vacuum with respect to linear polarization, the reflection from the rear face of the slab is a purely nonlinear effect. The bending of the lower branch in the figure as the front face of the slab is crossed, is likewise a purely nonlinear effect. 

Given that we essentially have weakly reflecting nonlinear mirrors at the two ends of the slab, one might wonder why we in figure \ref{fig8}, and also in figure \ref{fig11}, don't see higher order internal reflections from the left boundary. After all, the time window is large enough to include such reflections. 

The reason why we don't see such higher order reflections is because the reflectivity of the nonlinear mirrors defining the  boundaries of the slab is too low. The incoming left source field is of order one, in the scaled variables we are using. From the color scale in figure \ref{fig8} we see that the first order reflected field coming from the right boundary of the slab has a size of order $10^{-3}$. The second order reflection from the left boundary should then be expected to be of order $10^{-6}$. As one can see from the color scale for figure \ref{fig8}, the second order reflection is simply too weak to appear in the picture.

At second glance we see that figure display a profoundly acasual process. This is because {\it both} the upper and lower branches in figure \ref{fig8} come from the left moving field. If this fact is taken at face value we have a situation where two pulses, corresponding to the two branches in the figure, originate at the rear face of the slab. One pulse, the upper branch, travels forward in time, whereas the second pulse, corresponding to the lower branch, travels backwards in time. Our method appears to have converged to a acasual solution to the scattering problem. In section 4 of this paper we will present a strong argument for causality of the process displayed in figure \ref{fig8}. The apparent acausality of the process comes from a faulty interpretation of the picture, not a fault in our iteration method.

\begin{figure}[h!]
    \centering
    \includegraphics[width=0.8\linewidth]{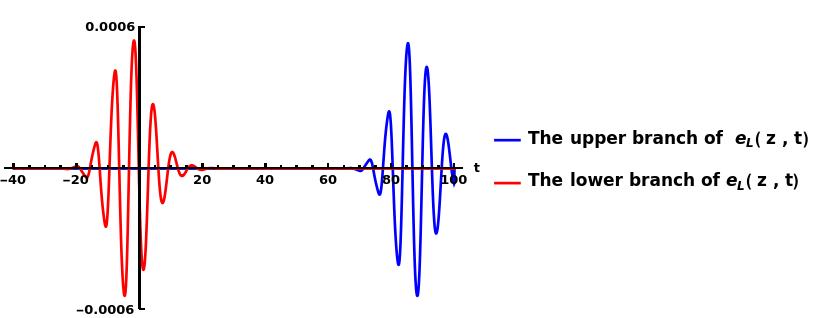}
    \caption{The real part of the right moving electric field as a function of time, taken at $z=5$ from figure \ref{fig8}.}
    \label{fig9}
\end{figure}

\FloatBarrier
In figure \ref{fig9} we show the temporal shape, at $z=5$,  of the two pulses represented by the lower and upper branches in the space-time picture from figure \ref{fig8}. The two pulses appear to be quite similar, but different too. It is easy to verify that the two pulses are in fact identical except for a phase shift of exactly $\pi$. Incidentally, this is exactly what we would expect if the first, causal interpretation of figure \ref{fig8} given above, is the correct one.

\FloatBarrier

\subsection{A long slab}
In this sub section we will consider the case of a slab of length equal to $150$ wavelengths. For the scaling we are using this corresponds to a slab of length approximately equal to $329 \mu m$.  In order to resolve the spectrum for this case we have used $8000$ discrete spectral points. 

Whether or not a causal or acausal explanation of figures \ref{fig6} and \ref{fig8} is the correct one, we  don't expect the space-time picture of the right and left moving fields to look very different from what they did for the short slab. From the two next figures \ref{fig10} and \ref{fig11}, we see that this is indeed the case.

\begin{figure}[h!]
    \centering
    \includegraphics[width=0.4\linewidth]{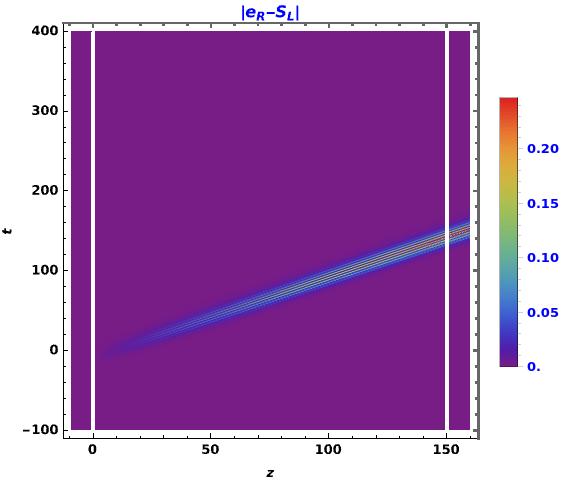}
    \caption{Space-time picture of the amplitude of the right moving part of the electric field with the left source subtracted. The white vertical lines defines the extent of a slab of length 150 wavelengths. Zero padding is $\omega_m=18$}
    \label{fig10}
\end{figure}

\begin{figure}[h!]
    \centering
    \includegraphics[width=0.4\linewidth]{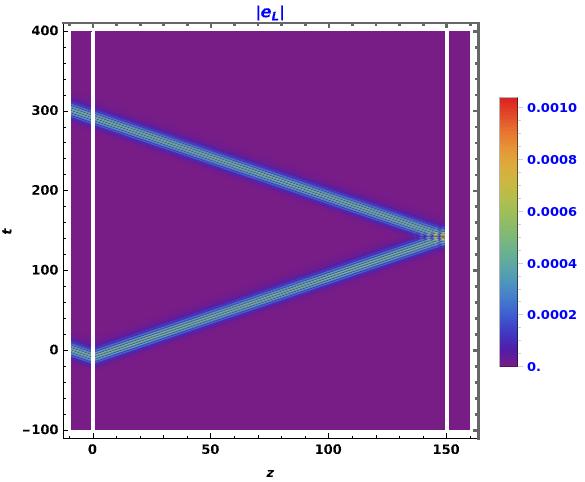}
    \caption{Space-time picture of the amplitude of the left moving part of the electric field. The white vertical lines defines the extent of a slab of length 150 wavelengths. Zero padding is $\omega_m=18$}
    \label{fig11}
\end{figure}

\begin{figure}
    \centering
    \includegraphics[width=0.6\linewidth]{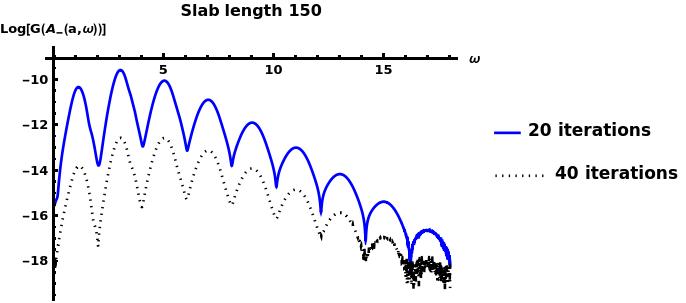}
    \caption{Logarithm of the remainder, $G(A_-(a,\omega))$, after 20 iterations, solid blue, and 40 iterations, dotted black, of the map (\ref{FixpointIteration}) for  a slab of length 150 wavelengths.  Zero padding $\omega_m=18$ }
    \label{fig12}
\end{figure}

The main point of this section is whether or not, and to which accuracy, our fix point iteration converge, for this longer slab, to the prospective solutions for the scattering problem. As figure \ref{fig12} show, it converge to a solution that after 40 iterations satisfy the exact equation (\ref{GMappingProblem}) to more than twelve digits of accuracy.

\FloatBarrier

\section{Fixpoint iterations create causal fields}
In the previous section we have seen that if the interpretation of $A_-(z,\omega)$, as the spectral amplitude of left moving waves, is correct, then our solution method leads to unacceptable acasual reflected fields. Here we will argue that it is not our method that is at fault, but rather the interpretation of $A_-(z,\omega)$. We will do this by applying our method to an exactly solvable case where we can fully see  what is actually going on. 

\begin{figure}[h!]
    \centering
    \includegraphics[width=0.7\linewidth]{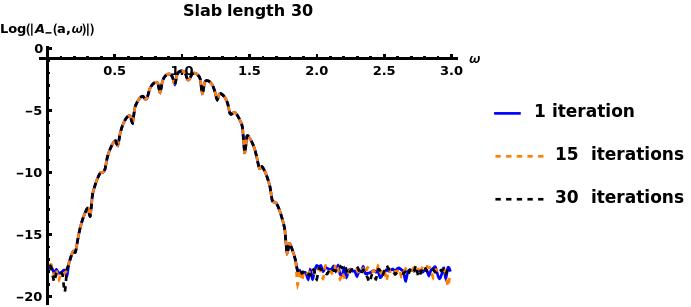}
    \caption{The spectrum for the linear system after 1 , 15 and 30 iterations of the map (\ref{FixpointIteration}) for a slab of length 30 wavelengths. Zero padding $\omega_m=3$}
    \label{fig13}
\end{figure}

We have all through this paper talked about the function $p(z,t)$ as representing the nonlinear part of the polarization inside the slab. Our method is however designed without putting any conditions on the actual form of this function. The method acts in the same way whatever the function is. 
Here we will take advantage of this by assuming that $p(z,t)$ is a {\it linear} function of the electric field of the form
\begin{align}
    p(z,t)=e(z,t).
\end{align}
Applying our fix point iteration to this linear case, for a slab of length 30 wave lengths, we find that the spectrum, displayed in figure \ref{fig13}, contains no new spectral peaks beyond the one given by the left source. This is what one would expect from what is a linear scattering problem. The fast oscillations on top of the spectrum is similar to the ones we saw for the nonlinear case treated in the previous section. The oscillation is caused by the finite length of the slab. They are slower than for the spectra in the previous section simply because the slab here is shorter.

\begin{figure}[h!]
    \centering
    \includegraphics[width=0.7\linewidth]{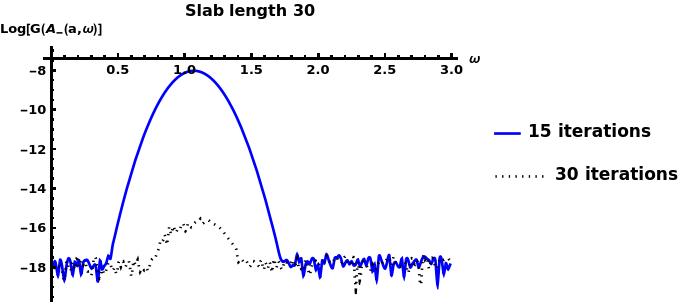}
    \caption{A logarithmic plot of the residual of the exact equation (\ref{GMappingProblem}) for a slab of length 30 wavelengths.  Zero padding $\omega_m=3$}
    \label{fig14}
\end{figure}

In figure \ref{fig14} we show the residual of the exact equation for this case. Our numerical solution satisfy the exact equation to machine precision.

\begin{figure}[h!]
    \centering
    \includegraphics[width=0.4\linewidth]{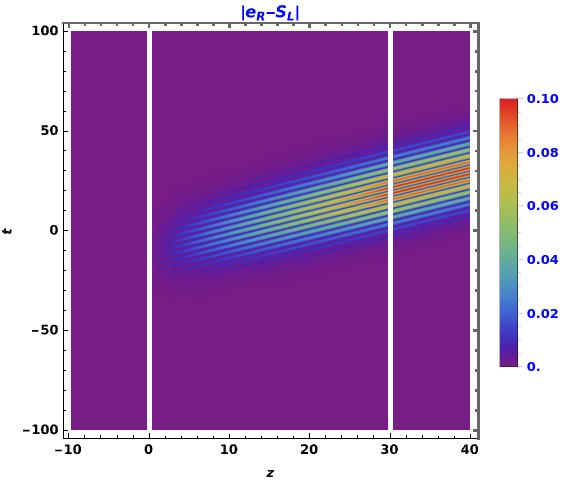}
    \caption{Space-time picture of the amplitude of the right moving part of the electric field inside the slab.  The white vertical lines defines the extent of a slab of length 30 wavelengths. Zero padding is $\omega_m=3$ }
    \label{fig17}
\end{figure}

\begin{figure}[h!]
    \centering
    \includegraphics[width=0.4\linewidth]{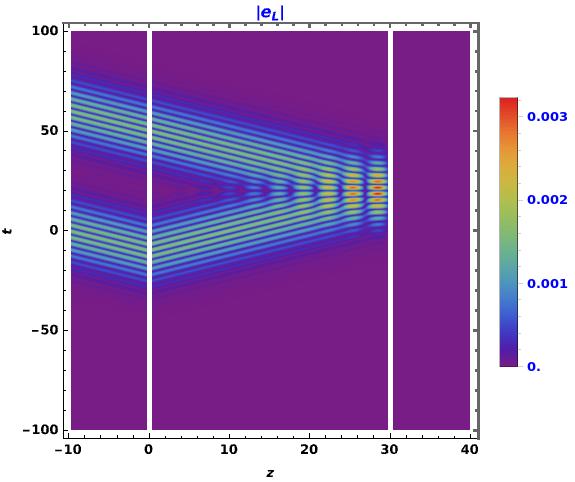}
    \caption{Space-time picture of the amplitude of the left moving part of the electric field inside the slab. The white vertical lines defines the extent of a slab of length 30 wavelengths. Zero padding is $\omega_m=3$}
    \label{fig18}
\end{figure}

The space-time figures \ref{fig17} and \ref{fig18} show the exact same apparent acausality as for the nonlinear case discussed in the previous section

For this case we can get full insight into the apparent acausality since linearity of the BPPE equations inside the slab makes them exactly solvable. The resulting expressions are fairly complicated and not very illuminating. However, if we take into account the smallness of the dimensionless parameter $\epsilon$ in the BPPE equations, we find for $\omega>0$, to leading order in $\epsilon$, the following much simpler expressions for $A_+(z,\omega)$ and $A_-(z,\omega)$
\begin{align}
    A_+(z,\omega)&=(1+\frac{\epsilon}{2} i\omega (z-a))S_L(\omega),\nonumber\\
    A_-(z,\omega)&=\frac{\epsilon}{4}(e^{2 i \omega b}-e^{2 i \omega z})S_L(\omega).
\end{align}
Here we have assumed that the right source is absent. This is the same assumptions as the one we used for the nonlinear slab in the previous section. 

Recall from the expansions (\ref{eFieldCausalityForm}) that the electric field inside a purely nonlinear slab is given by 
\begin{align}
    e(z,t)&=\int_{0}^{\infty}d\omega\;\left(\omega A_+(z,\omega)e^{i\omega\left( z-t\right)}+\omega A_-(z,\omega)e^{-i\omega\left(z+t\right)}\right)+(*).\nonumber\\
\end{align}
Given this, the left moving part of the electric field, $e_L(z,t)$ is , according to the interpretation of the spectral amplitudes given in section \ref{interpretation}, given by the formula
\begin{align}
    e_L(z,t)&=\int_{0}^{\infty}d\omega\;\omega A_-(z,\omega)e^{-i\omega\left( z+ t\right)}+(*)\nonumber\\
    &\approx \frac{\epsilon}{4}\int_{0}^{\infty}d\omega\;\omega S_L(\omega)(e^{2 i \omega b}-e^{2 i \omega z})e^{-i\omega\left( z+
    t\right)}+(*)\nonumber\\
    &=\frac{\epsilon}{4}\int_{0}^{\infty}d\omega\;\omega S_L(\omega)e^{2 i \omega b}e^{-i\omega(z+t)}-\frac{\epsilon}{4}\int_{0}^{\infty}d\omega\;\omega S_L(\omega)e^{i\omega(z-t)}+(*).
\end{align}
We observe that in fact, the "left moving" part of $e(z,t)$, $e_L(z,t)$,  consists of a left moving part 
\begin{align}
\frac{\epsilon}{4}\int_{-\infty}^{\infty}d\omega\;\omega S_L(\omega)e^{2 i \omega b}e^{-i\omega(z+t)}\label{LeftMoving},
\end{align}
and also a {\it right} moving part 
\begin{align}
    -\frac{\epsilon}{4}\int_{-\infty}^{\infty}d\omega\;\omega S_L(\omega)e^{i\omega(z-t)}\label{RightMoving}.
\end{align}
The left moving part (\ref{LeftMoving}), of $e_L(z,t)$, corresponds to the upper, causal, branch of the electric field in figure (\ref{fig18}), whereas the right moving part (\ref{RightMoving}),  corresponds to the lower, seemingly acasual, branch in the same figure. Thus,  the lower branch actually corresponds to a right moving part of the electric field, and is thus also a causal branch. There is no acasual branch, only a faulty  interpretation of the spectral amplitude $A_-(z,\omega)$. We conjecture that the same applied to the nonlinear case from the previous section.

Since we now realize that our designated left moving part of the electric field, $e_L$,  actually contains both a right moving and a left moving part, a peculiar feature of figures \ref{fig7} and \ref{fig11}, can be understood. The peculiar feature is that to the right of the slab, the field vanish completely. The reason for this is that to the right of our slab, where the light-matter interaction is purely linear, our defined left moving part of the electric field is exactly zero since there is no source of left moving waves in the region to the right of the slab.

Given this, a reader might then question why we did not rather restrict our plots to the full electric field, which is a simple sum of our left moving field $e_L$, and right moving field $e_R$.  The reason for this is that the size of the left moving field is much smaller than the right moving one. In a plot like that, only our right moving field $e_R$ would show up. 

In order to get a better physical insight into our scattering solutions it is possible to magnify $e_L$ by a factor of $100$, and then add the right moving part, $e_R$.  We still subtract the left source since it will otherwise end up dominating the picture completely. Since any reflection from the cavity walls is a purely nonlinear effect, the source pass more or less unchanged through the cavity. Only a very small part of the source engage in nontrivial  nonlinear reflections and transmissions generated by the cavity.

\begin{figure}[H]
    \centering
    \includegraphics[width=0.4\linewidth]{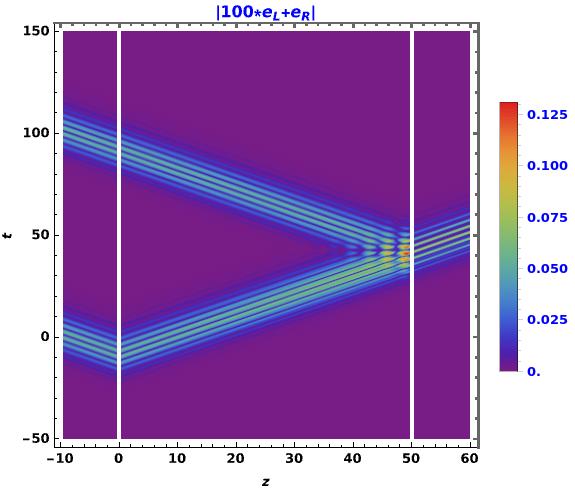}
    \caption{Space-time picture of the quantity $100e_L+e_R$ . The white vertical lines defines the extent of a slab of length 50 wavelengths. Zero padding is $\omega_m=18$}
    \label{fig86}
\end{figure}
This figure is fully physical and describe a situation where a source to the left of the slab generate a weak reflected field at the nonlinear mirror defining the left boundary of the slab. Inside the slab nonlinear interaction between the source field and the material in the slab generate a weak right moving electric field component. This nonlinearly generated field component is too weak to produce any significant reflection at the nonlinear mirror defining the right boundary of the slab. It thus essentially pass unchanged through the right boundary of the slab. This is what we see in figure \ref{fig86}.

However, the source is still as powerful as it was when it hit the left boundary, and therefore produce a reflected field at the nonlinear mirror defining the right boundary of the slab. This reflected field is so weak that it don't really engage with the nonlinear material in the slab and thus  propagate from the right boundary of the slab back to  the left boundary in a purely linear way. Being so weak, this reflected field pass through the nonlinear mirror defining the left boundary of the slab without any detectable reflection. We note that because the nonlinear mirrors defining the boundaries of the slab are so weak, the  amplitudes of the left moving reflected field generated at the to boundaries should be expected to be very close in size. This is why the size of the upper and lower branches of the reflected field to the left of the slab in figures \ref{fig7} and \ref{fig11}, appears to be the same.

%\begin{figure}[h!]
%    \centering
%    \includegraphics[width=0.6\linewidth]{Picture43.jpg}
%    \caption{The right moving electric field at $z=5$.}
%    \label{fig15}
%\end{figure}

%\begin{figure}[h!]
%    \centering
%    \includegraphics[width=0.6\linewidth]{Picture44.jpg}
%    \caption{The right moving electric field at $z=5$.}
%    \label{fig16}
%\end{figure}

\section{Method for generating exact solutions to the nonlinear scattering problem}

In this section we will not do the detailed derivation of our method for generating exact solutions to the basic nonlinear mapping problem (\ref{GMappingProblem}), these details are placed in appendix B. What we will do here is to create enough of a bird eyes view of the method so that one can follow the detailed derivation in appendix B, and also can generalize the method to other optical scattering problems beyond the simple application discussed in this paper.

We start by solving the BPPE system (\ref{ApAmEquations}) over the interval $a<z<b$ for values at $z=a$ which depends in a specific way  on a given right source $S_R(\omega)$, and another spectral function $\psi(\omega)$, that can be freely specified for positive frequencies. Let us denote the resulting solution of the BPPE system by $B_+(z,\omega)$ and $B_-(z,\omega)$.

We now flip the pair of functions $(B_+(z,\omega),B_-(z,\omega))$  and then  essentially reflect the flipped pair around the midpoint of the slab. Let us denote the resulting pair of functions by $(A_+(z,\omega),A_-(z,\omega))$.

Because of the reflection, what was the computed  values of the pair  $(B_+(z,\omega),B_-(z,\omega))$ at the {\it end} of the slab $z=b$, becomes a pair of function defined at the {\it start} of the slab at $z=a$. The first component of this pair becomes, by definition,  the spectral amplitude for the left source of the BPPE system for  $(A_+(z,\omega),A_-(z,\omega))$, and the second component becomes the spectral amplitude for the reflection from the slab, assuming that the right source is the given one, $S_R(\omega)$. In appendix B we prove that this last sentence is a true statement.

Note that these numerically constructed left sources do not have simple spectra, like the ones we have used earlier in this paper, and  are probably hard to design experimentally, but this is not really their purpose. Their purpose is to test the validity and accuracy of our fix point iteration method, or, for that matter,  any other method for solving the nonlinear mapping problem (\ref{GMappingProblem}).
 
In the three figures \ref{fig19}, \ref{fig20}, and \ref{fig21} below, we illustrate the theory described in this section, and fully derived in appendix B,  by showing that our iteration method converge to the exact reflection spectrum for the case when $\psi(\omega)$ in (\ref{BpBm}) is the same function as the one from (\ref{ALeftSource}).

\begin{figure}[h!]
    \centering
    \includegraphics[width=0.7
\linewidth]{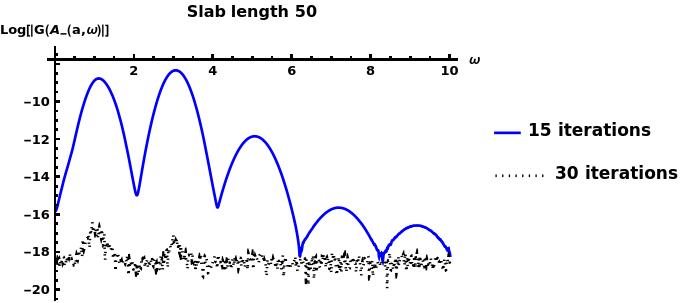}
    \caption{Logarithm of the residual, $G(A_-(a,\omega))$, after 15 iterations, solid blue, and 30 iterations, dotted black, of the map (\ref{FixpointIteration}) for  a slab of length 50 wavelengths.  Zero padding $\omega_m=18$ }
    \label{fig19}
\end{figure}

\begin{figure}[h!]
    \centering
    \includegraphics[width=0.9\linewidth]{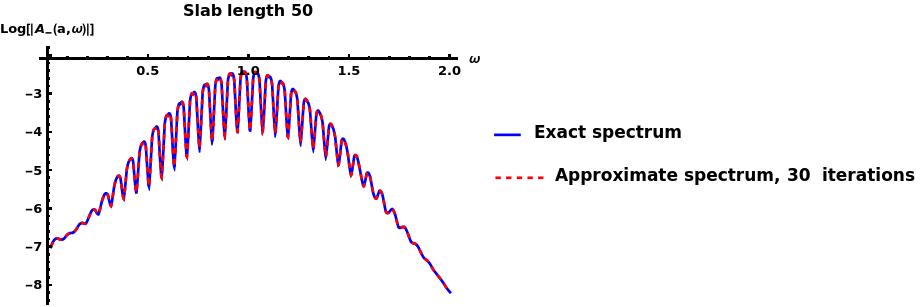}
    \caption{A logarithmic plot, over the range $0<\omega<2$, of the exact spectrum, solid blue, and the approximate spectrum after 30 iterations, dotted red.  Zero padding $\omega_m=18$}
    \label{fig20}
\end{figure}

\begin{figure}[h!]
    \centering
    \includegraphics[width=0.9\linewidth]{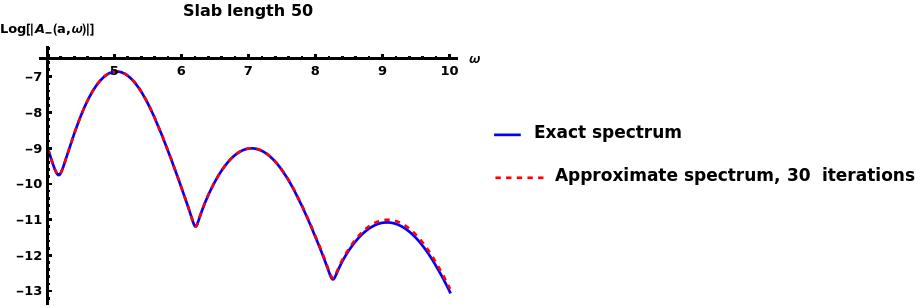}
    \caption{A logarithmic plot, over the range $4<\omega<10$, of the exact spectrum, solid blue, and the approximate spectrum after 30 iterations, dotted red.  Zero padding $\omega_m=18$}
    \label{fig21}
\end{figure}
\FloatBarrier

\section{Summary}
In this paper we have introduced a method for reformulating  optical scattering problems, linear or nonlinear,  into the solution of a mapping problem. For the detailed description of the method given is this paper to be directly applicable there are no specific constraints on the material properties of the scattering object, but there are geometrical constraints on the shape of the scattering object, which must be  a simple slab as described in the beginning of section 2 of this paper. In section 2 we both derive the mapping problem $G[A_-]=S_R$, and how the mapping problem can be turned into a fix point problem for a map which is a reduction and redefinition of our original map $G$.

We next show that for a specific type of nonlinear material response, which  is a sum of a fast electronic vibrational response, and slow molecular vibrational response, the fix point problem can be solved, in the numerical sense, by iteration of the reduced map.

We find that the solution found by iteration displays an apparent acausality. This apparent acausality is shown to be caused by a faulty interpretation of the spectral amplitudes. This is done by investigating a simplified situation, which  shows the same type of apparent acausality, and for which  the fix point problem for the reduced map is exactly solvable. 

In the final section of the paper we address the question of how we can ensure that iteration of our reduced map converge to the actual solution of the scattering problem. This question  have two different components. The first one is the question wether or not the iteration converges, and the second one is if it actually converges to the right thing. In the field of nonlinear optical  scattering rarely can either of the two components of our question be fully answered in the mathematical sense.

The first  component of our question we address in section 3 by showing that after a fairly small number of iterations the quantity $\norm{G[A_-]-S_R}$ stay numerically close to zero as the iteration proceed. This is what commonly done in applied science, where exact analytical results are few and far between, and where numerical approximations are basically all we have.

The second  component of our question can only be answered by knowing what the right solution to the nonliner scattering problem is. Knowing this in an analytical sense is almost always out of reach. The time honored way to conundrum is to solve the same scattering problem using different and independent methods, which are invariably also numerical in nature. In the current  paper we don't do this. What we rather do is construct a solution, also numerical, to our mapping problem, without relying on an iteration procedure. The construction is detailed in appendix B, and in section 5 is used to verify that iterating the reduced map does indeed converge to the right thing.

Our method can with small modifications be generalized to compound slabs consisting of several contiguous simple slab with varying linear and nonlinear material responses. These slabs can be bounded in the transverse direction and have a nontrivial geometric structure in that direction\cite{Miro4}. In fact our method can be applied to calculate reflection and transmission for any structure to which the well known UPPE method applies in the reflection less limit\cite{Miro3}.

It should also be clear that our method is not tied optical scattering, but can also be applied to the scattering of electromagnetic waves in general, or indeed to scattering of waves in general, including acoustic waves pressure waves, ocean internal and surface waves.

What we have not done in this paper is to compare our method to other methods for optical scattering, in domains of application where our method overlaps with these other methods. It is in particular an urgent problem to compare our method to FDTD for scattering problems where they both applies.

Another urgent issue which we have not discussed in our paper is to compare, with respect to accuracy and efficiency, our iteration method for the reduced map to methods directly  solving the nonlinear mapping problem $G[A_-]=S_R$, using the large body of methods known for these kind of problems. 

\section{Acknowledgment}
The author is thankful for support from the Department of mathematics and statistics at the Arctic University of Norway, from the Arizona Center for Mathematical Sciences at the University of Arizona, and for the support from the Air Force Office for Scientific Research under Grant No. FA9550-19-1-0032. The author are also thankful for insightful discussions and constructive critique of the ideas appearing in this paper from members of the Arizona Center for Mathematical Sciences located in the Wyant College of Optical Sciences, Arizona, USA. The author would in particular thank ACMS members, Jerry Moloney,  Miro Kolesik,  and former PhD student Andrew Hofstrand, which  has been involved in, and supportive of, the development and implementation of earlier versions of BPPE. 

%\section{Appendix}
%\appendix

\section*{Appendix A\label{appendixA}}
\setcounter{equation}{0}
\setcounter{section}{1}
\renewcommand{\theequation}{\Alph{section}.\arabic{equation}} 

In this appendix we do the details with respect to scaling of the model equations. 

We only need to consider a system of the form 
\begin{align}
    -\partial_z e + \mu_0 \partial_t h &= 0,\label{eq1} \\
   \partial_z h - \epsilon_0 \partial_t (e + p_L) &=  \epsilon_0\partial_t p.\label{eq2}
\end{align}
By letting $p_L=p_1, p_2, p_3$, in front of the slab, inside the slab, and behind the slab, and letting $p=0$ in front of the slab and behind the slab, we recover the model equations from the text.

We start by introduce arbitrary scales for all dependent and independent variables ion the equation. At this point, the scales are unknown.
\begin{align}
t&=\bar{t} t'\nonumber\\
z&=\bar{z} z' \nonumber\\
e&=\bar{e} e'\nonumber\\
h&=\bar{h} h'\nonumber\\
p&=\bar{p} p'
\end{align}
Here $e'$, $h'$, and $p'$ are functions of of dimensionless variables $z',t'$,
\begin{align}
e'(z', t') &= \frac{e(\bar{z} z', \bar{t} t')}{\bar{e}}\nonumber\\
h'(z', t') &= \frac{h(\bar{z} z', \bar{t} t')}{\bar{h}}\nonumber\\
p'(z', t') &= \frac{p(\bar{z} z', \bar{t} t')}{\bar{p}}
\end{align}
Given this, we evidently have
\begin{equation}
\partial_{t'} e' = \partial_{t'} \frac{e(\bar{z} z', \bar{t} t')}{\bar{e}} = \frac{\bar{t}}{\bar{e}} \partial_t e(z, t). 
\end{equation}
Treating derivative $e$ with respect to $z$ in the same way, we have
\begin{align}
\partial_t e &= \frac{\bar{e}}{\bar{t}} \partial_{t'} e' \nonumber \\
\partial_z e &= \frac{\bar{e}}{\bar{z}} \partial_{z'} e'.
\end{align}
Similar formulas hold for $h$ and $p$
\begin{align}
\partial_t h &= \frac{\bar{h}}{\bar{t}} \partial_{t'} h', \nonumber \\
\partial_z h &= \frac{\bar{h}}{\bar{z}} \partial_{z'} h',\nonumber\\
\partial_t p &= \frac{\bar{p}}{\bar{t}} \partial_{t'} p'.
\end{align}

Inserting these expressions into equation (\ref{eq1}) we get
\begin{align}
-\frac{\bar{e}}{\bar{z}} \partial_{z'} e' + \frac{\mu_0 \bar{h}}{\bar{t}} \partial_{t'} h' &= 0 \nonumber\\
&\Downarrow\nonumber\\
-\partial_{z'} e' + \frac{\mu_0 \bar{z} \bar{h}}{\bar{t} \bar{e}} \partial_{t'} h' &= 0 \nonumber \\
&\Downarrow \nonumber\\
-\partial_{z'} e' + \varepsilon_1 \partial_{t'} h' &= 0,\label{DimensionlessEq1}
\end{align}
where we have introduced the dimensionless quantity
\begin{equation}
\varepsilon_1 = \frac{\mu_0 \bar{z} \bar{h}}{\bar{t} \bar{e}} \label{eps1}
\end{equation}
From equation (42) we get
\begin{align}
\frac{\bar{h}}{\bar{z}} \partial_{z'} h' + \frac{\varepsilon_0}{\bar{t}} \partial_{t'} \left( \bar{e} e' + p_L\right)  &= \frac{\epsilon_0\bar{p}}{\bar{t}} \partial_{t'} p'\nonumber\\
&\Downarrow \nonumber\\\partial_{z'} h' - \frac{\varepsilon_0 \bar{e} \bar{z}}{\bar{h} \bar{t}} \partial_{t'} \left( e' + \frac{p_L}{\bar{e}}  \right) &= \frac{\epsilon_0\bar{z}\bar{p}}{\bar{h}\bar{t}} \partial_{t'} p'\nonumber\\
&\Downarrow\nonumber \\\partial_{z'} h' - \epsilon_2 \partial_{t'} \left( e' + \frac{p_L}{\bar{e}} \right) &= \epsilon_4 \partial_{t'} p',
\end{align}
where we have  define two  additional dimensionless quantities
\begin{align}
    \epsilon_2&=\frac{\varepsilon_0 \bar{e} \bar{z}}{\bar{h} \bar{t}},\\
    \epsilon_4&=\frac{\epsilon_0\bar{z}\bar{p}}{\bar{h}\bar{t}}.
\end{align}
For the linear polarization term we have
\begin{align}
    \frac{p_L}{\bar{e}}(z,t) &= \frac{1}{\bar{e}}\int_{-\infty}^{t} ds \, \chi_L(s - t) e(z, s)\nonumber\\
    &=\bar{t}\;\overline{\chi_L}\int_{-\infty}^{t'} ds' \, \frac{\chi_L(\bar{t}(t'-s'))}{\overline{\chi_{L}}} \frac{e(\bar{z} z',\bar{t},s')}{\bar{e}}\nonumber\\
    &=\epsilon_3\int_{-\infty}^{t'} ds' \, \chi'_L(s' - t') e'(z', s'),\label{DimesionlessDispersion}
\end{align}
where  $\overline{\chi_{L}}$ is a dimensional constant measuring the size of $\chi_L$, and where 

\begin{align}
    \chi'_L(\xi)&=\frac{\chi_L(\xi)}{\overline{\chi_{L}}},\nonumber\\
    \epsilon_3&=\bar{t}\;\overline{\chi_L}.\label{chiAndEps4}
\end{align}
From the above derivations and definitions we conclude that the dimensionless version of equation (\ref{eq1}), is
\begin{align}
     \partial_z' h' - \epsilon_2 \partial_t' (e' + \epsilon_3 p'_L) &=  \epsilon_4\partial_t' p',\nonumber
\end{align}
where 
\begin{align}
    p'_L=\int_{-\infty}^{t'} ds' \, \chi'_L(s' - t') e'(z', s').\label{DimensionlessEq2}
\end{align}

The dimensionless equations (\ref{DimensionlessEq1}) and (\ref{DimensionlessEq2}) is simplified by choosing scales such that $\epsilon_1=\epsilon_2=\epsilon_3=1$. It is easy to see that these conditions are satisfied if 
\begin{align}
    \bar{z}&=c\bar{t},\nonumber\\
    \bar{e}&=\sqrt{Z_0 \bar{I}},\nonumber\\
     \bar{h}&=\sqrt{\frac{\bar{I}}{Z_0}},\nonumber\\
    \overline{\chi_L}&=\frac{1}{\bar{t}},\label{Scales}
\end{align}
where $c$ is the speed of light,  $Z_0=\sqrt{\frac{\mu_0}{\epsilon_0}}$ is the vacuum impedance, and $\bar{I}=\bar{e}\bar{h}$ is equal to the intensity for a mode of the electromagnetic field of amplitude $(\bar{h},\bar{e})$ traveling along the $z$-axis.

Note that the scales $\bar{I}$, $\bar{t}$, and $\bar{p}$, are still arbitrary at this point, and that the sole remaining dimensionless parameter in the model is
\begin{align}
     \epsilon\equiv\epsilon_4=\frac{\epsilon_0c\sqrt{Z_0}\bar{p}}{\sqrt{\bar{I}}}.
\end{align}

In this paper we assume that the formulas for the Fourier transform looks the same in scaled and unscaled variables. For this to be the case there need to be a particular relation between the scales for the Fourier transform of the function and the scales for the function and time. There also needs to be a particular relation between the scales for frequency and time.
We introduce scales for $f$, $\hat{f}$, and $\omega$ through the identities
\begin{align}
  f&=\bar{f} f',\nonumber\\
  \hat{f}&=\bar{\hat{f}}\hat{f}',\nonumber\\
  \omega&=\bar{\omega}\omega'.\label{FourierScales}
\end{align}
Inserting these expressions into the first of the equations defining the Fourier transform (\ref{MyFourier}), we have 
\begin{align}
    \bar{\hat{f}}\frac{\hat{f}(\bar{\omega}\omega')}{\bar{\hat{f}}}&=\bar{f}\bar{t}\frac{1}{\sqrt{2\pi}}\int_{-\infty}^{\infty}dt'\;\frac{f(\bar{t}t')}{\bar{f}}e^{i\bar{\omega}\bar{t}\omega' t'},\nonumber\\
    &\Downarrow\nonumber\\
    \hat{f}'(\omega')&=\frac{\bar{f}\bar{t}}{ \bar{\hat{f}}}\frac{1}{\sqrt{2\pi}}\int_{-\infty}^{\infty}dt'\;f'(t')e^{i\bar{\omega}\bar{t}\omega' t'},\label{DFourier1}  
\end{align}
where we have defined the dimensionless functions 
\begin{align}
    f'(t')&=\frac{f(\bar{t}t')}{\bar{f}},\nonumber\\
    \hat{f}'(\omega')&=\frac{\hat{f}(\bar{\omega}\omega')}{\bar{\hat{f}}}.\label{DimensionlessFourierTransformPair}
\end{align}
From (\ref{DFourier1}) it is evident that the Fourier transform relation connecting the unscaled variables $f$, and $\hat{f}$ are the same as the one connecting the scaled Fourier transform variables $f'$, and $\hat{f}'$ only if 
\begin{align}
    \bar{\hat{f}}&=\bar{f}\bar{t},\nonumber\\
    \bar{\omega}&=\frac{1}{\bar{t}}.\label{FourierScales}
\end{align}
It is easy to check that if we assume the choice of scales (\ref{FourierScales}), then the second equation defining the Fourier transform also looks the same in scaled and unscaled variables.

In section 3 of the main text we introduce a specific choice for the nonlinear polarization
\begin{align}
    p(z,t)&=\left((1 - \theta) p_K(z,t) + \theta p_R(z,t)\right).\label{NonlinearPolarization}
\end{align}
Here  $\theta$ is a dimensionless constant measuring the relative size of the two components $p_K$ and $p_R$, which are given by 

\begin{align}
    p_K(z, t) &=  e(z, t) \int_{-\infty}^{t} ds \, \chi_K(s - t) e^2(z, s), \\
    p_R(z, t) &= e(z, t) \int_{-\infty}^{t} ds \, \chi_R(s - t) e^2(z, s).
\end{align}
Note that at this point all variables are dimensional quantities, scales have not been introduced yet.
Introducing the scales in the expression for $p_K$ we have
\begin{align}
   p_K(z, t) &=  \bar{e}^3\;\overline{\chi}\bar{t}\;\;\frac{e(\bar{z}z', \bar{t}t')}{\bar{e}} \int_{-\infty}^{t'} ds' \, \frac{\chi_K(\bar{t}(s' - t'))}{\overline{\chi}} \left(\frac{e(\bar{z}z', \bar{t}s')}{\bar{e}}\right)^2, \nonumber\\  
 &=\left(\frac{\overline{\chi}}{\overline{\chi_L}}\right)\bar{e}^3\;\overline{\chi_L}\bar{t}\;\;\frac{e(\bar{z}z', \bar{t}t')}{\bar{e}} \int_{-\infty}^{t'} ds' \, \frac{\chi_K(\bar{t}(s' - t'))}{\overline{\chi}} \left(\frac{e(\bar{z}z', \bar{t}s')}{\bar{e}}\right)^2,\nonumber \\  
 &=\left(\frac{\overline{\chi}}{\overline{\chi_L}}\right)\bar{e}^3\;\;e'(z',t')\int_{-\infty}^{t'} ds' \,\chi'_K(s'-t') e'(z',s')^2,\nonumber\\
 &=\eta\bar{e}^3\;\;e'(z',t')\int_{-\infty}^{t'} ds' \,\chi'_K(s'-t') e'(z',s')^2,
\end{align}
where by definition 
\begin{align}
    \chi'_K(\xi)&=\frac{\chi_K(\bar{t}\xi)}{\overline{\chi}},
\end{align}
and where
\begin{align}
    \eta=\left(\frac{\overline{\chi}}{\overline{\chi_L}}\right),
\end{align}
is a dimensional constant measuring the relative size of the nonlinear and linear polarization.

Treating the expression for the other component of the nonlinear polarization, $p_R$, in the same way, we get from (\ref{NonlinearPolarization}) that
\begin{align}
    p(z,t)=\bar{p}\left((1 - \theta) p'_K(z',t') + \theta p'_R(z',t')\right),
\end{align}
where $z=\bar{z}z',t=\bar{t}t'$ and where
\begin{align}
    \bar{p}&=\eta\bar{e}^3,\\
    p'_K(z',t')&=e'(z',t')\int_{-\infty}^{t'} ds' \,\chi'_K(s'-t') e'(z',s')^2,\\
    p'_R(z',t')&=e'(z',t')\int_{-\infty}^{t'} ds' \,\chi'_R(s'-t') e'(z',s')^2,
\end{align}

\section*{Appendix B\label{appendixB}}
\setcounter{equation}{0}
\setcounter{section}{2}
\renewcommand{\theequation}{\Alph{section}.\arabic{equation}}

  The goal is to find a pair of functions $\phi_{+}(\omega),\phi_{-}(\omega)$ such that 
\begin{align}
P \circ M_{23}(\omega) \circ U(a, b) \circ M_{12}(\omega) 
\begin{pmatrix}
\phi_+(\omega) \\
\phi_-(\omega)
\end{pmatrix}=S_R(\omega).\label{Goal}
\end{align}
Given that the goal has been achieved, we now simply define
\begin{align}
    S_L(\omega)&=\phi_+(\omega),\nonumber\\
    A_-(a,\omega)&=\phi_-(\omega),\nonumber
\end{align}
and conclude that we have a solution, $A_-(a,\omega)$ to the nonlinear mapping problem (\ref{GMappingProblem})
\begin{align}
    G(A_-(a,\omega))=S_R(\omega),\nonumber
\end{align}
corresponding to the given left and right sources, $S_L(\omega)$ and $S_R(\omega)$.

In order to achieve this goal, we will start by rewriting  the BPPE equations inside the slab into the form
\begin{align}
    2 i\beta_2\partial_z A_+(z, \omega) e^{i\beta_2 z}&= -\epsilon\omega \hat{p}(z,\omega), \nonumber\\
     -2 i\beta_2\partial_z A_-(z, \omega) e^{-i\beta_2 z}&= -\epsilon\omega \hat{p}(z,\omega), \quad a < z < b.\label{ApAmEquations}
\end{align}
At this point we need to be a little more precise about the nature of the function $\hat{p}(z,\omega)$ appearing in (\ref{ApAmEquations}). In general, $\hat{p}$ is a map whose value at each point $(z,\omega)$, is a functional of $e(z,t)$.  The expression (\ref{NonlinearPolarization}),  we used in section  three, is an example of this.  Since the electric field itself is a linear functional of the form
\begin{align}
 e(z,t)&=\int_{-\infty}^{\infty}d\omega\;\hat{e}(z,\omega)e^{-i\omega t},\nonumber
\end{align}
where 
\begin{align}
    \hat{e}(z,\omega)=\omega A_+(z,\omega)e^{i\beta_2 z}+\omega A_-(z,\omega)e^{-i\beta_2 z},\nonumber
\end{align}
we can without loss of generality write  the function $\hat{p}(z,\omega)$ in the form
\begin{align}
    \hat{p}(z,\omega)=h(z,\omega)[\omega A_+(z,\omega)e^{i\beta_2 z}+\omega A_-(z,\omega)e^{-i\beta_2 z}].\nonumber
\end{align}
Here $h$ is a function whose value at a point $(z,\omega)$,   is a functional. The square brackets signify the action of this  functional on spectrum of the electric field.

We now proceed by defining a matrix $J$ of the form
\begin{align}
J=  \left(
\begin{array}
[c]{cc}
0 &e^{-i\beta_2 (a+b)}\\
e^{i\beta_2 (a+b)} & 0
\end{array}\right).\nonumber
\end{align}

\noindent Next, let  $B_{+}(z,\omega), B_{-}(z,\omega)$ be the unique solution to the BPPE system (\ref{ApAmEquations}) for $a<z<b$, corresponding to the  boundary conditions
\begin{align}
  \left(
\begin{array}
[c]{c}
B_{+}(a,\omega)\\
B_{-}(a,\omega)
\end{array}\right)&=J^{-1}M_{32}^{-1}(\omega)\left(
\begin{array}
[c]{c}
\psi(\omega)\\
S_R(\omega)
\end{array}\right),\label{BpBm}
\end{align}
where $\psi(\omega)$ is a spectral density function that can be chosen freely on half of the full spectral range $-\infty<\omega<\infty$, for example the positive range $0<\omega<\infty$. The reason for this restriction is the reality condition $\psi^*(-\omega)=S_R(\omega)$, which must be imposed in order to ensure that the pair of spectral densities $\psi(\omega), S_R(\omega)$ define a physical electromagnetic field in the region $z>b$, through the mode expansion formulas (\ref{ModeExpansions}).

Using this reality condition, and the special structure of the matrices $M_{32}$ and $J$, it is easy to verify that the spectral densities $B_{+}(a,\omega), B_{-}(a,\omega)$ in (\ref{BpBm}) also satisfy the reality constraints. Furthermore, since the reality constraint is conserved by the BPPE system, we can conclude that the two density functions  $B_{+}(z,\omega), B_{-}(z,\omega)$ define a physical electromagnetic field inside the slab.

Given $B_{+}(z,\omega), B_{-}(z,\omega)$ , we define  two density functions $A_{+}(z,\omega),A_{-}(z,\omega)$ using
\begin{align}
  \left(
\begin{array}
[c]{c}
A_{+}(z,\omega)\\
A_{-}(z,\omega)
\end{array}\right)&=\left(
\begin{array}
[c]{c}
B_{-}(a+b-z,\omega)e^{-i\beta_2 (a+b)}\\
B_{+}(a+b-z,\omega)e^{i\beta_2 (a+b)}
\end{array}\right)=J\left(
\begin{array}
[c]{c}
B_{+}(a+b-z,\omega)\\
B_{-}(a+b-z,\omega)
\end{array}\right).\label{ApAmDefinition}
\end{align}
Writing $\xi=a+b-z$, and observing that $a<\xi<b$,  we now have
\begin{align}
2 i\beta_2\partial_z A_{+}(z,\omega)e^{i\beta_2 z}&=-2 i\beta_2\partial_{\xi} B_{-}(\xi,\omega)e^{-i\beta_2\xi}\nonumber\\
&=-\epsilon\omega h(z,\omega)[B_{+}(\xi,\omega) e^{i\beta_2 \xi}+B_{-}(\xi,\omega) e^{-i\beta_2 \xi}]\nonumber\\
&=-\epsilon\omega h(z,\omega)[B_{+}(\xi,\omega) e^{i\beta_2(a+b)}e^{-i\beta_2 z}+B_{-}(\xi,\omega)(\xi,\omega) e^{-i\beta_2(a+b)} e^{i\beta_2 z}]\nonumber\\
&=-\epsilon\omega h(z,\omega)[A_{-}(z,\omega) e^{-i\beta_2 z}+A_{+}(z,\omega) e^{i\beta_2 z}]\nonumber\\
&=-\epsilon\omega h(z,\omega)[A_{+}(z,\omega) e^{i\beta_2 z}+A_{-}(z,\omega) e^{-i\beta_2 z}]=-\epsilon\omega\hat{p}(z,\omega)\nonumber.
\end{align}
For the second equation of the BPPE system we get in a similar way
\begin{align}
-2 i\beta_2\partial_z A_{-}(z,\omega)e^{-i\beta_2 z}&=2 i\beta_2\partial_{\xi} B_{+}(\xi,\omega)e^{i\beta_2\xi}\nonumber\\
&=-\epsilon\omega\hat{p}(z,\omega).\nonumber
\end{align}
Thus the spectral amplitudes $A_{+},A_{-}$  are the solutions to the BPPE equations for $a<z<b$.

Finally, we define a pair of functions $\phi_{+}(\omega),\phi_{-}(\omega)$ in the following way
\begin{align}
\left(
\begin{array}
[c]{c}
\phi_{+}(\omega)\\
\phi_{-}(\omega)
\end{array}\right)=M_{21}^{-1}(a)  \left(
\begin{array}
[c]{c}
A_{+}(a,\omega)\\
A_{-}(a,\omega)
\end{array}\right).\nonumber
\end{align}
Then we have achieved our goal, because the constructed pair of functions, $A_{+},A_{-}$, are  the unique solutions to BPPE inside the interval $a<z<b$ corresponding to the boundary condition
\begin{align}
 \left(
\begin{array}
[c]{c}
A_{+}(a,\omega)\\
A_{-}(a,\omega)
\end{array}\right)=M_{12}(\omega)\left(
\begin{array}
[c]{c}
\phi_{+}(\omega)\\
\phi_{-}(\omega)
\end{array}\right),\label{FirstPart}
\end{align}
and at the right boundary we have
\begin{align}
P\circ  M_{23}(\omega) \left(
\begin{array}
[c]{c}
A_{+}(b,\omega)\\
A_{-}(b,\omega)
\end{array}\right)&=P\circ M_{23}(\omega)\circ J\left(
\begin{array}
[c]{c}
B_{+}(a,\omega)\\
B_{-}(a,\omega)
\end{array}\right)\nonumber\\
&=P\circ M_{23}(\omega)JJ^{-1}M_{23}^{-1}(\omega)\left(
\begin{array}
[c]{c}
\psi(\omega)\\
S_R(\omega)
\end{array}\right)= P\left(
\begin{array}
[c]{c}
\psi(\omega)\\
S_R(\omega)
\end{array}\right)=S_R(\omega).\label{SecondPart}
\end{align}
Putting together (\ref{FirstPart}), (\ref{SecondPart}), and the fact that the spectral functions defined in (\ref{ApAmDefinition}) satisfy the BPPE system, we conclude that (\ref{Goal}) holds.

\end{document}